\documentclass[a4paper,12pt,headsepline]{scrartcl}

\usepackage{graphicx}

\usepackage{natbib}

\usepackage{comment}

\usepackage{xtab}

\usepackage{lineno}

\usepackage{makecell}

\usepackage{pdflscape}

\usepackage{subcaption}

\usepackage{geometry}
\usepackage[hyphens,obeyspaces,spaces]{url}

\usepackage{setspace}

\usepackage{xcolor}

\usepackage{colortbl}
\definecolor{coolgrey}{rgb}{0.80, 0.80, 0.80}
\definecolor{battleshipgrey}{rgb}{0.90, 0.90, 0.90}

\usepackage{longtable}

\usepackage{outlines}

\usepackage{ngerman}

\usepackage[utf8]{inputenc}

\usepackage{gensymb}

\usepackage{amsmath} 

\usepackage{amssymb}

\usepackage{hyperref} 
\hypersetup{colorlinks=true, citecolor=black, linkcolor=black, urlcolor=black}
\usepackage{pdfpages}

\begin{document}
\selectlanguage{english}

\clearpage{\pagestyle{empty}\cleardoublepage}

\thispagestyle{empty}
\doublespacing
\begin{center}
{\large \textbf{Actors in Multi-Sector Transitions -}}\\ 
\vspace*{1mm}
{\large \textbf{Discourse Analysis on Hydrogen in Germany}}\\ 
\vspace{1.0cm}
%
\begin{normalsize}
Nils Ohlendorf\textsuperscript{1,2,*} 
Meike Löhr\textsuperscript{3} 
Jochen Markard\textsuperscript{4,5}\\
\end{normalsize}
\end{center}
\singlespacing
\vspace{1.0cm}
%
\begin{footnotesize}
\textsuperscript{1} Technische Universität Berlin, Department of Economics of Climate Change, Straße des 17. Juni 145, 10623 Berlin, Germany \\
\textsuperscript{2} Mercator Research Institute on Global Commons and Climate Change, Torgauer Staße 12-15, 10829 Berlin, Germany \\
\textsuperscript{3} Carl von Ossietzky University Oldenburg, Institute of Social Sciences, Oldenburg, Germany \\
\textsuperscript{4} Swiss Federal Institute of Technology in Zürich, Department of Management, Technology and Economics, Zürich, Switzerland \\
\textsuperscript{5} Zurich University of Applied Sciences, Institute of Sustainable Development, Winterthur, Switzerland
\end{footnotesize}
\\
\vspace{0.5cm}
\\
\textsuperscript{*}Corresponding author: Nils Ohlendorf, Torgauer Staße 12-15, 10829 Berlin, Germany
\\
Mail: ohlendorf\_nils@hotmail.com
\\
\vspace{0.5cm}
\\
\textbf{Acknowledgements:}
\\
Nils Ohlendorf acknowledges funding by the German Federal Ministry of Education and Research (PEGASOS, funding code 01LA1826C). 
Meike Löhr (meike.loehr@uni-oldenburg.de) acknowledges funding by the German Research Foundation (grant number 316848319) in the context of the Emmy Noether-project ``Regional energy transitions as a social process'' (REENEA).
Jochen Markard (jmarkard@ethz.ch) acknowledges funding from the Norwegian Research Council (Conflicting Transition Pathways for Deep Decarbonization, grant number 295062/E20) and from the Swiss Federal Office of Energy (SWEET program, PATHFNDR consortium).

We thank participants and discussants at the International Sustainability Transitions Conference (IST 2021),
5th International Conference on Public Policy (ICPP5),
the Sustainable Technologies Colloquium at the ETH Zürich,
and members of the Climate and Development group at the MCC Berlin for their helpful comments and suggestions. 
\\
\vspace{0.5cm}
\\
\textbf{Declaration of Interest:}
\\
The authors declare that they have no known competing financial interests or personal relationships that could have appeared to influence the work reported in this paper.
\\
\vspace{0.5cm}
\\
\textbf{Please cite as:} Ohlendorf, N., Löhr, M., Markard, J. (2023). Actors in multi-sector transitions-discourse analysis on hydrogen in Germany. Environmental Innovation and Societal Transitions, 47, 100692.
\\

\clearpage{\pagestyle{empty}\cleardoublepage}

\thispagestyle{empty}
\doublespacing
\begin{center}
{\large \textbf{Actors in Multi-Sector Transitions -}}\\ 
\vspace*{1mm}
{\large \textbf{Discourse Analysis on Hydrogen in Germany}}\\ 
\vspace{1.0cm}
%
\begin{normalsize}
\end{normalsize}
\end{center}
\singlespacing
\textbf{Abstract:}
\\
With net-zero emission goals, low-carbon transitions enter a new phase of development, leading to new challenges for policymaking and research. Multiple transitions unfold in parallel across different sectors. This involves a broad range of technologies, while actors engage in increasingly complex discourses. Here, we study the discourses on hydrogen in Germany. Based on the analysis of 179 newspaper articles from 2016 to 2020, we find that a diverse set of actors, including many industry incumbents, speak favorably about hydrogen, emphasizing economic opportunities and its relevance for the energy transition, whereas skeptics highlight its low energy efficiency and expected scarcity. With the help of discourse network analysis, we identify three emerging conflicts around the use, production, and import of hydrogen. We explain these conflicts and the widespread support of incumbents with a conceptual framework that captures the complex interplay of sectoral contexts, specific technologies and actor interests.
\\
\vspace{1.0cm}

\clearpage{\pagestyle{empty}\cleardoublepage}

\setlength{\columnsep}{25pt}

\doublespacing

\section{Introduction}\label{sec:introduction}

As more and more countries and businesses are making pledges to reduce their greenhouse gas emissions to net-zero until mid-century \citep{hohne_wave_2021},
the transition toward low-carbon energy systems is entering a new phase of development. It is not sufficient any more to pursue incremental changes, or to focus on selected sectors, such as electricity or transport. Instead, societies need to cut, or compensate for all greenhouse gas emissions across all sectors.

This ‘net-zero phase’ of the energy transition imposes new challenges. One challenge is the simultaneous involvement of multiple socio-technical systems, or sectors, and various technologies within and across these multiple systems. Another challenge is the large number of actors with different backgrounds and interests. Transitions research has begun to address some of these challenges, highlighting the increasing complexity if transitions comprise multiple sectors or multiple technologies \citep{papachristos_system_2013, rosenbloom_engaging_2020, andersen_multi-technology_2020}.
Despite some progress, we still lack empirical studies and conceptual frameworks to analyze transitions that involve multiple systems \citep{rosenbloom_engaging_2020}.

We address this gap with a study on the emerging field of hydrogen, a research case that is at the heart of the latest phase of the net-zero energy transition \citep{van_renssen_hydrogen_2020}.
Hydrogen is currently pushed by policymakers as an alternative energy carrier for a broad range of applications, including difficult-to-decarbonize industries (DDI), such as steel or aviation \citep{davis_net-zero_2018}.
As of 2021, 17 countries have already adopted hydrogen strategies, while another 20 are under development \citep{iea_global_2021}.
We select Germany as a leading industrialized country with many large incumbent firms in different industries. The German government actively promotes hydrogen as an essential part of its net-zero strategy.

We ask how actors talk about hydrogen, and whether and why they support or oppose to it. We also address how incumbent actors may be affected by the sectoral context they are operating in. We approach these questions by analyzing the discourses on hydrogen in Germany in leading nationwide newspapers. Our analysis starts in 2016 and covers the years leading up to the end of 2020. The final dataset comprises 179 newspaper articles of five newspapers, quotes of 139 actors, and 30 storylines. Discourse analysis \citep{hajer_doing_2006} can be a particularly insightful tool for innovations at early stages of development when actors try to create meaning and common understanding, shape categories, or influence policies \citep{rosenbloom_framing_2016}. 
We also employ discourse network analysis (DNA) \citep{victor_discourse_2017} on a subset of storylines related to three emerging conflicts.

We find a widespread, partly even enthusiastic support for hydrogen among incumbent actors, while environmental NGOs and some think tanks adopt more skeptical positions. Considering that incumbents have often resisted transformative change \citep{hess_sustainability_2014, smink_keeping_2015} this might seem surprising at first. Our findings add to more recent studies that add more nuance to the role of incumbents in transitions \citep{turnheim_forever_2020}. We also identify three specific lines of conflict, namely about in i) which sectors hydrogen should be used, ii) which production methods are desirable, and iii) potential risks and benefits of hydrogen imports.

We also contribute by developing a simple conceptual framework to study the involvement of multiple sectors, multiple technologies and multiple actors in discourses around emerging sustainability innovations. The framework considers the sectoral context, sector-specific technologies, and actor interests, with a special focus on incumbent actors. We then apply our conceptual framework to the case of hydrogen.

The paper proceeds as follows; 
Section \ref{sec:theory} describes the theoretical background, 
Section \ref{sec:methodology} explains our approach and methodology. 
Section \ref{sec:results} presents the results. 
Section \ref{sec:discussion} applies and discusses the framework. 
Section \ref{sec:concluding_remarks} concludes.

\section{Theoretical background}\label{sec:theory}

Our study is rooted in the literature on sustainability transitions, which studies fundamental changes in socio-technical systems as a response to grand sustainability challenges such as climate change \citep{markard_sustainability_2012, kohler_agenda_2019}.\footnote{In the following, we use the terms 'system' and 'sector' interchangeably.} 
We understand the net-zero energy transition as a complex, long-term transformation process that involves multiple transitions in different socio-technical systems. This novel setting is conceptually demanding because we are confronted with i) multiple systems in different transition stages, ii) complex interactions of multiple technologies, and iii) a broad range of actors, with incumbent actors playing a very active role. Below, we briefly review these three issues. We also explain the merits of discourse analysis, and conclude with a conceptual framework that guides our reasoning, and facilitates the interpretation of the results.

\subsection{Interaction of multiple systems}\label{sec:interaction_systems}

Originally, socio-technical transitions have been depicted in a rather straightforward way: innovations emerge in niches, improve and diffuse over time, until eventually, one innovation has matured sufficiently to replace established practices and technologies, thereby fundamentally transforming a socio-technical system \citep{geels_technological_2002, markard_sustainability_2012}. 
In this view, a transition is primarily confined to a single system, which limits the number of innovations, technologies and actors involved. Later, scholars have shown that innovations may not just interact with one, but with multiple systems at once \citep{geels_typology_2007, raven_co-evolution_2007, konrad_multi-regime_2008, papachristos_system_2013}.
One example is biogas technology: it connects agriculture on the input side with different energy sectors on the output side \citep{sutherland_conceptualising_2015, markard_institutional_2016}. 
Another example are multi-purpose or general-purpose technologies that can be used for a broad range of applications \citep{dolata_technological_2009}.

In the case of the net-zero energy transition, the situation is even more complex because many sectors and industries transform at the same time and there are several interdependencies. For example, the transition towards renewable energies in the electricity sector enables the decarbonization of transport with electric mobility \citep{zhang_role_2020}.
Similar interdependencies arise for hydrogen, which might emerge as a viable option to tackle difficult-to-decarbonize industries \citep{davis_net-zero_2018, bataille_physical_2020},
while low-carbon electricity or other inputs are required to produce hydrogen in the first place.
The transitions literature is only beginning to grasp the complexity of multi-transition dynamics and multi-system interactions \citep{rosenbloom_engaging_2020}.

\subsection{Interaction of multiple technologies}

Next to multi-system dynamics, the net-zero energy transition is also characterized by an interaction of multiple technologies \citep{andersen_multi-technology_2020}. 

Transitions research has already described various forms of technology interaction and the conditions, under which technologies compete or complement each other \citep{sanden_framework_2011, markard_analysis_2016}.
In the electricity sector, the transition towards renewables is already in full swing \citep{mitchell_momentum_2016}, 
and in transport and heating, technologies that rely on low-carbon electricity diffuse in many places \citep{iea_global_2021, iea_heat_2021}. 
Hydrogen is expected to emerge as an alternative low-carbon energy carrier, either in addition to electricity, or as the primary alternative to fossil fuels. As a consequence, hydrogen-based technologies may both complement, or compete with existing and alternative low-carbon technologies. What complicates the situation is that some hydrogen technologies and infrastructures may reach across sectors, which means that system and technology interaction overlap.

\subsection{Actors in transitions}\label{sec_actors}

Actors such as firms, industry associations, think tanks, NGOs or policymakers play a crucial role in transitions \citep{farla_sustainability_2012}.
While they pursue different, possibly conflicting interests and strategies, they deploy various kinds of resources, forge networks and engage in institutional or transition work \citep{binz_thorny_2016, lohr_institutional_2022}.
Overall, they seek to shape policies \citep{musiolik_networks_2012, wesseling_car_2014},
and influence transitions through their strategy choices \citep{lohr_facing_2022}.

Of particular interest are incumbent actors, especially when they are economically well equipped and exert political influence \citep{turnheim_forever_2020}.
Some incumbents control critical resources (e.g., access to specific customers or suppliers), which is why they can be central for developing, or slowing down, new technologies and their diffusion \citep{rothaermel_complementary_2001, berggren_transition_2015}.
Many studies have found incumbent actors fighting against major technological or institutional changes to protect their established businesses \citep{hess_sustainability_2014, jacobsson_politics_2006, penna_multi-dimensional_2012, wesseling_car_2014, smink_keeping_2015}.

However, incumbent firms may also support or even drive transitions \citep{turnheim_forever_2020, lohr_energietransitionen_2020}. 
Transition scholars have shown pro-active strategies of incumbents in various sectors, including heavy vehicles \citep{berggren_transition_2015},
electrical engineering, 
automotive \citep{bergek_technological_2013}, 
or horticulture \citep{kishna_innovation_2017}. 
In these examples, incumbents support transitions when they can continue their established business \citep{berggren_transition_2015}, or when innovation builds on their existing business model \citep{bergek_technological_2013}. In this study, we also find a diverse and even leading role of incumbents. Section \ref{sec:discussion_categories} provides suggestions to explain their support (or resistance).

\subsection{Discourse analysis}\label{sec:DNA}

When actors talk about innovations or transitions more broadly, they convey meaning, shape categories and (co-)create expectations \citep{borup_sociology_2006, bakker_arenas_2011, roberts_politics_2018}.
Especially when they talk in public, they often pursue specific interests such as mobilizing policy support \citep{budde_expectations_2012, isoaho_politics_2020}
or influencing the perceived legitimacy \citep{geels_cultural_2011, rosenbloom_framing_2016}.
Public debates are thus inherently political. Through the analysis of argumentative structures, we can shed light on the underlying politics \citep{hajer_doing_2006}.

To analyze the public debate around hydrogen, we mobilize argumentative discourse analysis, which rests on the assumption that language, and the exchange of arguments, are key to understand the political nature of an issue, and to identify political conflicts \citep{hajer_decade_2005, brink_words_2006, hajer_doing_2006, isoaho_critical_2019, lowes_heating_2020}.
We understand discourse as ``\textit{as an ensemble of ideas, concepts and categories through which meaning is given to social and physical phenomena, and which is produced and reproduced through an identifiable set of practices.}'' \citep[p.67]{hajer_doing_2006}.
As a key element of our analysis, we create storylines, which are ``\textit{condensed statement[s] summarizing complex narratives, used by people as ’short hand’ in discussions}'' \citep[p.69]{hajer_doing_2006}.

Discourse analysis can be particularly useful to study technologies in early stages of development, when uncertainty is high \citep{binz_thorny_2016}, and pathways (including specific configurations and applications) are still to be shaped \citep{rosenbloom_framing_2016}. 
In early discourses, we expect that different groups of actors seek to influence the public debate, in our case, the formulation and implementation of the German hydrogen strategy.

In our paper, we use discourse analysis both qualitatively, and in a quantitative way using DNA. The latter allows us to systematically trace which actors make similar arguments \citep{leifeld_political_2012, victor_discourse_2017}. 
DNA has been applied to a broad range of issues including pension policy \citep{leifeld_reconceptualizing_2013}, 
nuclear and coal phase-out \citep{rinscheid_crisis_2015}, 
climate policy \citep{fisher_where_2013, kukkonen_international_2018}, 
energy transitions \citep{brugger_influence_2021}, or 
genetically modified organisms \citep{tosun_mobilization_2017}. 
DNA is compatible with Hajer’s conceptualization of discourse because it entails both a substantive dimension (arguments expressed through storylines), and a relational dimension in the form of actors sharing similar storylines \citep{victor_discourse_2017}.

\subsection{Conceptual framework}\label{sec:concept_framework}

This section seeks to integrate the different aspects we introduced above to explain how storylines form in a multi-sector, multi-technology setting. We built the framework from scratch, but aligned our analytical categories of actors, content and context to Rosenbloom et al.'s \citeyearpar{rosenbloom_framing_2016}, and transferred them to a setting that is characterized by multiple systems and technologies.

We distinguish between the 1) \textit{sectoral context} (with multiple sectors in different transition stages), 2) multiple \textit{sector-specific technologies} (with different socio-technical characteristics), and 3) strategic \textit{actor interests}. Moreover, we assume that all sectors are affected by broader societal and overarching technological changes. The underlying idea is that how actors talk about an innovation depends on the availability and progress of other innovations, and how the innovation relates to established technologies, infrastructures, business practices, and prior strategic commitments.

\begin{figure}[!htb]
\caption{Conceptual framework}\label{fig:concept_frame}
\includegraphics[width=\textwidth]{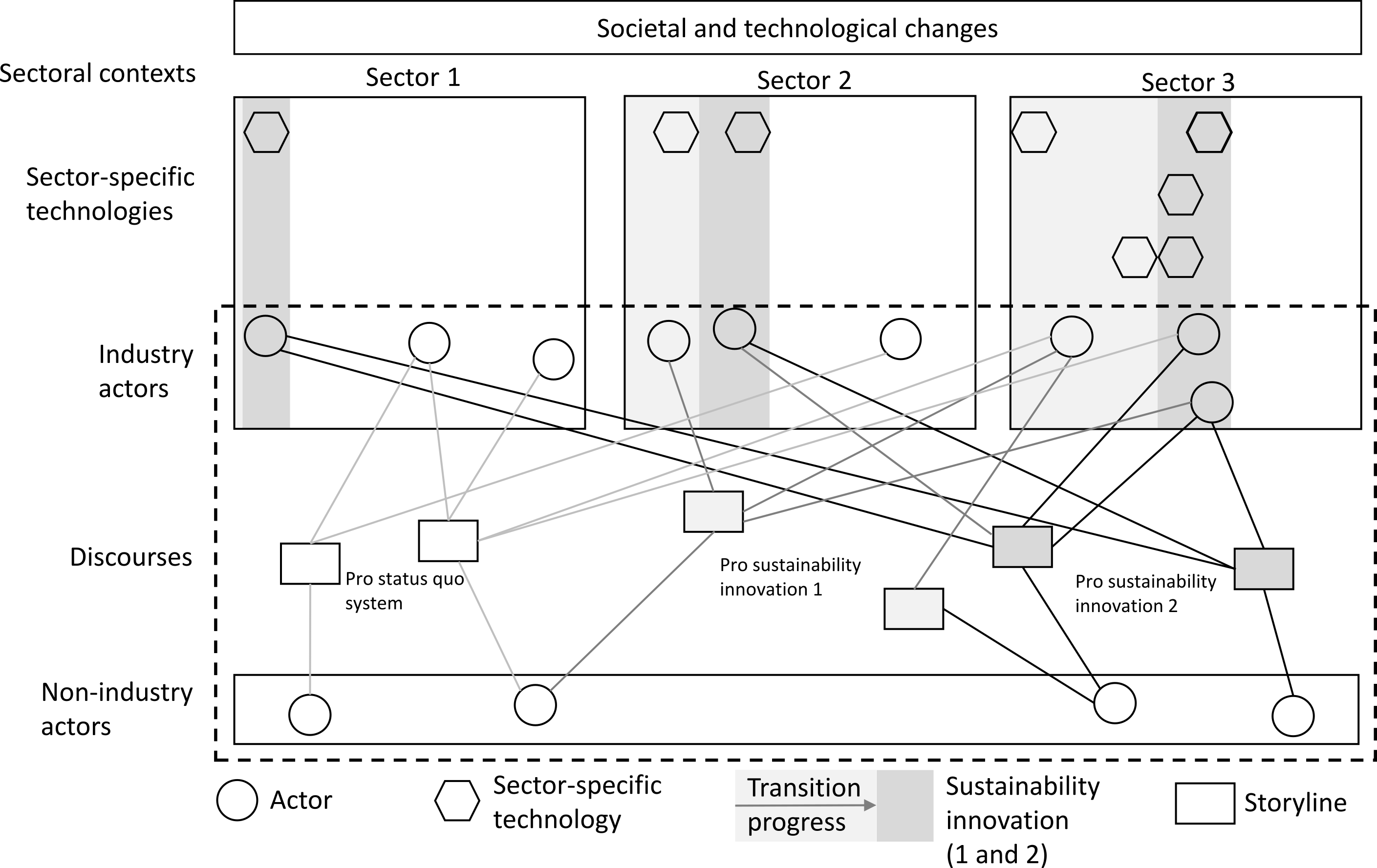}
\caption*{\footnotesize
The conceptual framework describes the behavior of actors in the public debate from three different sectors at different stages of transitions. Each sector has a status quo socio-technical configuration (white area), that is challenged by one or two emerging and competing sustainability innovations (grey areas) and associated sector-specific technologies (hexagons). The position of industry actors (circles on top) in relation to these areas reflect the commitment of actors to the respective technologies. All actors, including non-industry actors from politics, or the civil society, and from economic sectors that are not directly affected by the sustainability innovation (circles at bottom), participate in the discussion by uttering storylines (rectangles) that favor one of the three options. Moreover, we assume that all sectors are affected by broader societal and technological changes.}
\end{figure}

\autoref{fig:concept_frame} shows our generic framework, in which we distinguish i) three different sectors in different transition stages, ii) two sustainability innovations and one incumbent practice (‘status quo’), iii) industry actors in the sectors iv) non-industry actors, and v) discourses comprising of a broad set of storylines on both innovations and the incumbent practice. In our example, Sector 1 is at a very early sustainability transition stage and has only one sustainability innovation and sector-specific technology. In
the public debate, some industry actors already support the innovation, while others still promote the status-quo system. In Sector 2, the innovations have already made some progress (the white space for incumbent practices decreases) with two equally advanced competing technologies. The actors are divided between the options. Sector 3 has transitioned furthest and includes technologies at different stages of maturity. The actors have abandoned the status quo, but are divided between the different innovations. We apply the framework in Section \ref{sec:discussion}.

\section{Methodology}\label{sec:methodology}

The following sections describe our research case Germany, the selection of newspaper articles, the development of the coding scheme, the resulting storylines, the overall sample, and finally, our analytical approach.

\subsection{Hydrogen in Germany}

We select Germany as our research case for studying the emerging discussion on hydrogen. Germany has Europe’s largest economy, and the fourth largest nominal GDP on the globe \citep{imf_world_2021}.
Germany’s decisions about energy policies are important internationally, as they have already in the past spilled-over to other countries. For example, since the early 2000s, Germany’s \textit{Energiewende} demonstrated that a renewable energy transition is generally feasible, induced significant price reductions and efficiency improvements for renewables, and also promoted the expansion of renewables internationally \citep{quitzow_german_2016}.
Understanding the German energy transition is also particularly challenging and complex due to the large variety of established industries with its diverse set of actors.


Especially the goal to decarbonize the entire economy made hydrogen come to the fore of the German government. In early 2019, the German chancellor Angela Merkel announced the new goal of achieving the net-zero emissions goal until mid-century, in contrast to the previous goal of decreasing carbon emissions by 80-95\%. This paradigm shift implies decarbonizing the entire economy, including DDI sectors, such as the steel or chemical industry, or shipping and aviation \citep{davis_net-zero_2018, bataille_physical_2020}.
Other reasons driving the recent uptake of hydrogen are increasingly cheap renewable electricity, while increased prices of European Emission Allowances put additional pressure on energy intensive industries.

Germany adopted a national hydrogen strategy in June 2020 as part of an economic stimulus package in response to the Covid-19 crisis \citep{bmwi_national_2020}.
The strategy covers national and international projects on the generation, transport, distribution and use of hydrogen. Germany envisages becoming a large-scale hydrogen importer, also from emerging economies. The strategy explicitly focuses on green hydrogen from renewable energies, while blue hydrogen based on fossil fuels has been declared a temporary solution only.\footnote{Hydrogen can be produced in different ways, often distinguished by colors. For example, green hydrogen is produced from water via electrolysis using renewable electricity, while blue and grey hydrogen are obtained from natural gas, either using carbon-capture and storage to reduce carbon emissions (blue), or not (grey).}

\subsection{Article selection and coding}\label{sec:newspapers}

We investigate the public debate on hydrogen in Germany by analyzing articles of five leading newspapers comprising the 
\textit{Süddeutsche Zeitung}, 
\textit{Frankfurter Allgemeine Zeitung},
\textit{Die Welt},
\textit{Handelsblatt} and
\textit{taz}.
This selection covers five out of six most printed national daily newspapers in Germany with a circulation above 750.000 per day \citep{statista_uberregionale_2021},
and addresses the entire political spectrum from left (\textit{taz}) to mid-right (\textit{Die Welt}).
We select newspapers as they build an important platform for different actors to construct, vocalize and legitimize their envisaged societal role of hydrogen, and to thereby influence policymaking. To identify relevant newspaper articles, we developed a search query that covers a comprehensive selection of articles, while keeping the number of unrelated findings to a necessary minimum. The query includes articles with hydrogen mentioned at least once at the beginning, and four times in the main text.\footnote{The specific search parameters depend on the search options provided by the different databases used to search for the articles. These databases comprise Lexis 
(\textit{Die Welt}, \textit{Frankfurter Allgemeine Zeitung} and \textit{TAZ}),
WISO (\textit{Handelsblatt})
and the SZ archive (\textit{Süddeutsche Zeitung}).
The coding was done using the qualitative data analysis software MAXQDA.}
To ensure that we include the start of the recent discussion around hydrogen, we select articles starting with the year 2016 when the Paris Agreement was adopted. The search period ends with December 2020.\footnote{We are aware that hydrogen has been at the center of attention already around the year 2000 and several
times thereafter \citep{konrad_strategic_2012, budde_tentative_2019}.
Yet, the previously narrow focus on transport differs from the ongoing debate.}

The coding scheme was developed inductively and bottom-up in multiple iterations between the author team \citep{kuckartz_qualitative_2016}.
To familiarize with the topic, we complemented and triangulated our desk research via semi-structured interviews with nine hydrogen experts from NGOs, research institutes and a German ministry by mostly two or three authors in early 2021. To develop an initial coding scheme, each author independently read selected newspaper articles and proposed potential storylines. The selected articles were chosen to represent the entire range of discourses, and thus include long articles from all five newspapers with different topics and many direct quotes. The proposed storylines were subsequently discussed multiple times between the authors. In a next step, one author coded more than 10\% of the sample and thereby refined the initial coding scheme. All authors afterwards coded selected articles to compare their coding decisions, and to discuss and resolve potential ambiguities. One author then coded all articles using the final coding scheme.\footnote{The coding-tree in MAXQDA consists of short versions of each code. The long version, additional explanations, and coding examples are attached to coding memos.} 
All coded passages that contribute to the results in Section \ref{sec:res_conflicts} 
were checked by the entire author team to ensure the absence of coding errors. We only coded passages that agree to a storyline, and not such that oppose to it. Identified double codes arising from identical text passages within different articles were removed.

\subsection{The storylines}\label{sec:storylines}

This analysis includes 30 storylines, which we grouped into six topics. Topics 1-3 are about the deployment of hydrogen in general, and 4-6 refer to specific lines of conflict. Topics 1-3 address the role of hydrogen for climate change mitigation (1), economic considerations (2), and technical aspects (3) in a prospective renewable electricity system. In total, 9 storylines address these topics, while each respective topic is covered by three storylines. The remaining 21 storylines relate to three specific conflicts on the production method (4), hydrogen imports (5), and the use of hydrogen (6).
\autoref{tab:storyline_overview} shows the short and long version of the storylines. The storylines referring to the specific lines of conflict are aggregated and assigned to either one of two opposing positions. 
\autoref{tab:storyline_conflicts} shows the short versions of the disaggregated storylines in the specific conflicts.

\begin{landscape}
{
\def\sym#1{\ifmmode^{#1}\else\(^{#1}\)\fi}

\begin{table}

\caption{Storyline overview}\label{tab:storyline_overview}

\centering

\footnotesize
\onehalfspacing

\renewcommand{\arraystretch}{1}
\setlength{\tabcolsep}{5pt}

\begin{xtabular}{p{6cm}p{18cm}}
\hline
\rule{0pt}{3ex}Storyline short & Storyline long\\ 
\hline
\textbf{1. Climate change mitigation} (123) &\\
Important for energy transition  & Hydrogen is an important part of the energy transition. \\
Required for complete decarbonization  &Hydrogen is indispensable for achieving a complete decarbonization. Renewable electricity alone is insufficient. \\ 
Prioritize other climate mitigation options  & The focus on hydrogen risks to distract from other mitigation options (e.g. renewable deployment, efficiency, sufficiency) that should be prioritized. \\
\textbf{2. Economic considerations} (114) & \\
Economic opportunities & Hydrogen creates economic opportunities (e.g. technology exports, jobs, new value chains) for Germany's economy or individual companies. \\
Eventually cheap option   & Hydrogen will become a cheap energy carrier or at least profitable business case once cost reduction potentials are realized, or carbon costs increase further. \\
Generally scarce and expensive & Hydrogen will remain a scarce and expensive energy carrier, also in the future.\\
\textbf{3. Technical aspects} (71) & \\
Utilization of existing gas grid & Hydrogen allows utilizing the existing gas grid infrastructure. \\
Facilitates renewable integration  & Hydrogen facilitates the integration of (excess) renewable energies and stabilizes the electricity system.\\
Low energy efficiency  & Producing hydrogen is inefficient due to high energy losses. \\
\textbf{4. Use} (191) & \\
Wide use & Hydrogen should be used for private cars, domestic heating, blended into the natural gas grid, or generally applied widely. \\
Restricted use &  Hydrogen should primarily be used for difficult-to-decarbonize industries, or not for private cars, heating or be blended into the natural gas grid. \\
\textbf{5. Production method} (171) & \\
Non-green necessary & Non-green hydrogen is required for a transition period, or relying on green hydrogen alone is not possible, expensive, or risky. \\
Exclusively green &  Only green hydrogen should be used, or only green is carbon free, or CCS is contested, or non-green hydrogen prolongs using fossil fuels. \\
\textbf{6. Imports} (85) & \\
Imports beneficial &  Hydrogen imports will be comparatively cheap due to better wind and solar conditions, or exports will foster the economic development abroad. \\
Imports concerning &  Hydrogen imports lead to new energy dependency from other countries, or other countries may face environmental problems (water scarcity) or human rights violations. \\
\hline
\end{xtabular}
\caption*{\footnotesize 
The table groups each storyline under one of six topics and shows a short (1$^{st}$ column) and a long version (2$^{nd}$ column).
The number of coded passages for each topic is shown in brackets. Topics 4-6 aggregate several individual storylines; disaggregated short versions of these are shown in \autoref{tab:storyline_conflicts}.}
\end{table}
}

\end{landscape}

{
\def\sym#1{\ifmmode^{#1}\else\(^{#1}\)\fi}

\begin{table}

\caption{Emerging conflicts}\label{tab:storyline_conflicts}

\centering

\footnotesize
\onehalfspacing

 \begin{tabular}{lcr}
\hline
Enthusiastic: && Skeptical: \\
\hline
\rule{0pt}{3ex}\textbf{4. Use (191)} & & \\ 
Wide use (94): & $\longleftrightarrow$ & Restricted use (97): \\
\hspace{3mm} Private cars&& Not private cars \hspace{3mm}  \\
\hspace{3mm} Gas grid&& DDI priority \hspace{3mm} \\
\hspace{3mm} Heat&& Not heat \hspace{3mm} \\
\hspace{3mm} Wide application && Not gas grid \hspace{3mm} \\
\rule{0pt}{3ex}\textbf{5. Production method (171)} & & \\ 
Non-green necessary (88):  & $\longleftrightarrow$ &  Exclusively green (83): \\
\hspace{3mm} Transition and market creation&&       Explicitly green only \hspace{3mm} \\
\hspace{3mm} Consider various colors&        &          Only green carbon free \hspace{3mm} \\
\hspace{3mm} Only green not possible&&                       CCS is contested \hspace{3mm} \\
\hspace{3mm} Include grey&&                             Not green prolongs fossil fuel use \hspace{3mm} \\
\rule{0pt}{3ex}\textbf{6. Imports (85)} & & \\ 
Imports beneficial (55): & $\longleftrightarrow$ & Imports concerning (30): \\
\hspace{3mm} Using beneficial solar conditions&& Import dependency \hspace{3mm} \\
\hspace{3mm} Advantage for exporters&& Disadvantage for exporters \hspace{3mm} \\
\hspace{3mm} Potentially cheap imports && \\
\hline
\end{tabular}
\caption*{\footnotesize 
The table shows short versions of storyline related to the three emerging conflicts.
The frequency of coded passages for each aggregated storyline is shown in brackets,
sub-storylines are sorted descending by the frequency of coded passages.
Storylines on the left are enthusiastic about hydrogen, 
those on the right skeptical.}
\end{table}
}


The analysis only includes a selection of coded storylines that we consider most insightful to understand controversies in the public debate. It omits more detailed statements related to individual sectors or technology-specific aspects.\footnote{Statements on individual sectors are only included if they more generally address the conflict about where to use hydrogen.}
Excluded storylines address potential consequences from deploying hydrogen, consensual statements related to the conflicts, uncontested applications of hydrogen, and very rare storylines.

\subsection{The sample}\label{sec:sample}

The final sample comprises 179 newspaper articles with 614 coded passages by 139 actors. The initial search yielded 321 newspaper articles. Almost 4000 passages were coded. To condense the analysis to the most relevant aspects, we excluded several storylines (see previous section), removed false positive articles or articles without coded passages, duplicate codes by the same actor within single articles, and codes by journalists. Appendix \ref{appendix:search_query} provides further information on the search query, and more details about the sample of newspapers.

The actors covered by the analysis stem from a variety of fields and industrial sectors. To facilitate the overview, we aggregate individual actors to groups, namely \textit{Policymakers}, actors from the \textit{Transport, Gas and heat, Industry and Electricity sectors}, actors from \textit{Research and think tanks}, and \textit{NGOs}.\footnote{The
group \textit{Research and think tanks} also includes international organizations and consultancies. 
The majority of \textit{NGOs} focus on environmental topics.
The \textit{Industry} groups comprises potential users of hydrogen, but also producers.
The analysis omits individuals that cannot be assigned to any actor group.}
Within each industrial sector, the public debate is dominated by incumbents, while newcomers only play a minor role.

The number of coded passages between actor groups is very unequal. Policymakers are most salient, specifically different ministries, but also members of the EU Commission and political parties. Of industry actors, selected companies and industry associations from the automotive energy and gas sector are particularly visible (VW, BDI, RWE, Westenergy, Zukunfts Erdgas, FNB Gas, and Siemens). Other frequently occurring actors comprise specific NGOs (BUND, DUH, Klimaallianz) and research institutes (Max Planck Institute, Dena, Fraunhofer ISE).\footnote{In some instances, single quotes are cited in multiple articles. This mostly applies specifically to statements by ministers, and the CEO of the German car manufacturer Volkswagen}.
Appendix \ref{appendix:descr_actors} provides the descriptive statistics about discursive engagement of all actors. Appendix \ref{appendix:att_storylines} shows the temporal development of the public debate.

\subsection{Data analysis}\label{sec:data_analysis}

We analyze the sample in two steps using Stata. In a first step, we descriptively analyze the storylines that refer to using hydrogen in general by comparing the shares of each actor group. In a second step, we apply DNA to analyze the three specific lines of conflict regarding the use, production, and imports of hydrogen, based on codes of 21 storylines shown in \autoref{tab:storyline_conflicts}.

We restrict the DNA to the most active actors, only including actors with at least three coded passages.\footnote{Our inclusion criterion focuses on actors that are most visible in the overall public debate. We additionally created an actor congruence network that includes actors that are most visible in the conflicts (in contrast to the overall public debate) by tightening the inclusion criterion towards at least three coded passages within the three conflicts. The resulting network shows a qualitatively similar pattern, but only includes roughly half of the actors, with particularly many missing actors from the grey shaded area.}
The sample thereby reduces to 257 coded storylines by 63 actors. We create an actor congruence network to visualize individual actors using similar storylines. Links between actors are normalized to account for unequal numbers of coded storylines between actors, by dividing the edge weight with the average number of storylines \citep{victor_discourse_2017}.
\autoref{fig:disc_net} shows the resulting discourse network in detail. 
\autoref{fig:lines_of_conflict} shows the same discourse network, but with all edges highlighted that include at least one storyline in support of the respective conflict position. All figures are created with Stata and edited with Inkscape.

Data limitations require consideration when interpreting our findings. Gaining a comprehensive understanding of positions of most actors is impossible due to scant text passages with sufficient detail. The short analysis period, moreover, prevents analyzing dynamic changes of individual positions,
which thus remains subject for future research. We discuss further limitations in Section \ref{sec:limitations}.

\section{Results}\label{sec:results}

We identified six major topics in the discussion on hydrogen in Germany. Three topics address the deployment of hydrogen in general (Section \ref{sec:res_temp_dev}), and three address specific conflicts (Section \ref{sec:res_conflicts}).

\subsection{General topics: decarbonization, economic and technical aspects}\label{sec:res_temp_dev}

\begin{figure}[!htb]
\caption{General storylines}\label{fig:stat_general}
\includegraphics[width=\textwidth]{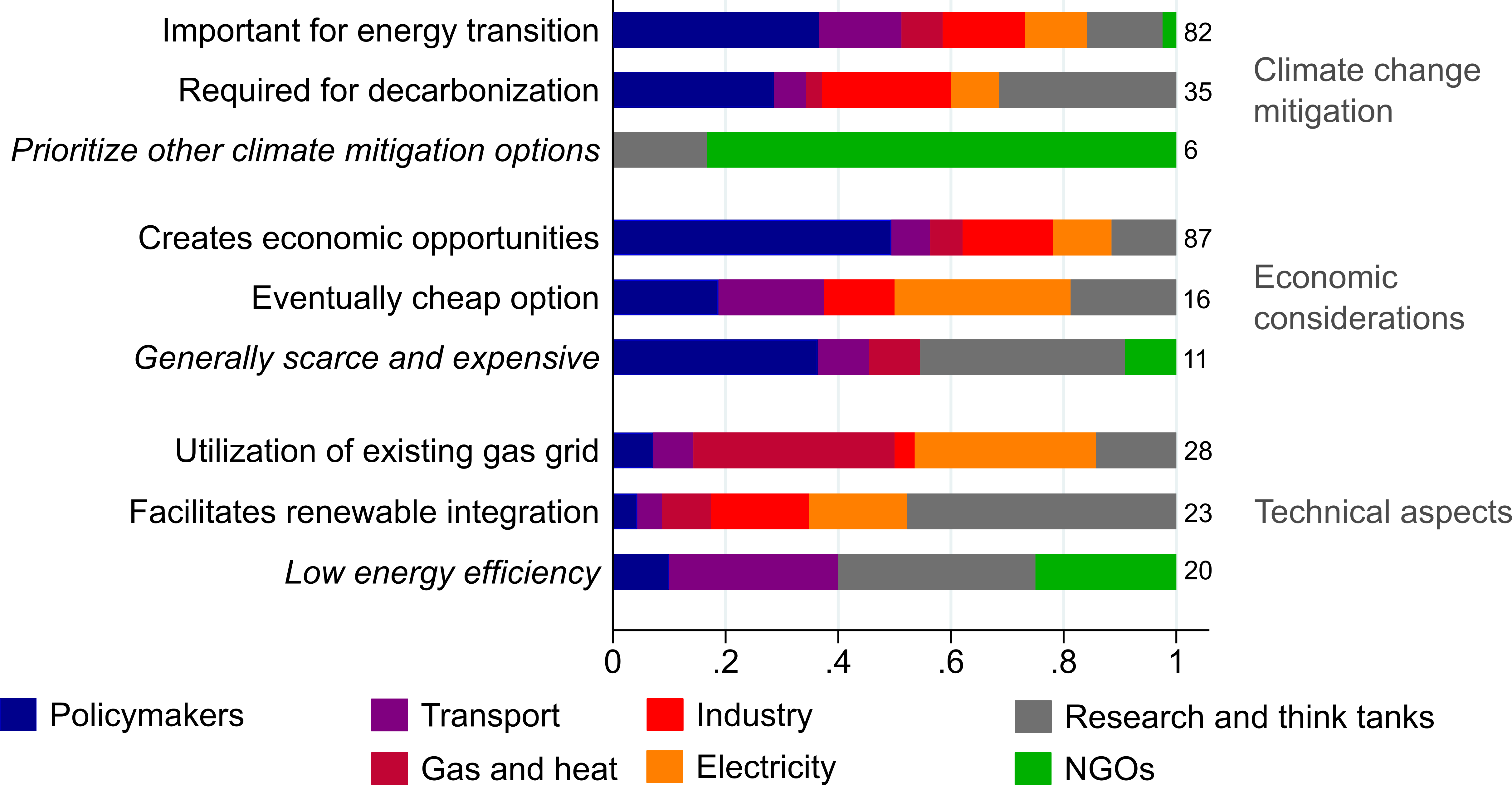}
\caption*{\footnotesize
The figure shows the share of general storylines referred to by each actor group. The nine storylines are grouped under three topics. The first two storylines of each topic are enthusiastic, the third takes a skeptical position (in italics). The frequency of coded passages is depicted next to the bars. Further information on the content of each storyline is provided in \autoref{tab:storyline_conflicts}.}
\end{figure}

Attention for hydrogen has significantly increased since 2019, and there is a lively debate between a broad range of actors from different sectors and institutional backgrounds. Interestingly, we find many incumbent firms from e.g., energy, transport, and industry. In general, there is widespread support that hydrogen is necessary to achieve net-zero, and that the development of hydrogen-based technologies creates economic opportunities. However, potential caveats receive attention as well.
Three overarching topics emerged: decarbonization, economic considerations, and technical aspects. Coincidentally, each topic entails two enthusiastic, and one skeptical storyline. 
\autoref{fig:stat_general} shows the relative shares of all nine storylines by actor group.

On the first topic, policymakers, industry actors and think tanks frequently mention that hydrogen is important for the energy transition, and required to reach net-zero. ``\textit{I am convinced that we cannot achieve the energy transition without gaseous energy sources. They are an indispensable part of the energy transition in the long term}'' (BMWi, 63).\footnote{The number corresponds to the list of newspaper articles in \autoref{tab:newspaper_articles}, Appendix \ref{appendix:search_query}.} 
NGOs do not necessarily disagree, but highlight that other climate mitigation options such as efficiency or sufficiency should be prioritized, or at least not be neglected. \begin{quote}
\textit{``At the moment, people like to pretend that there is an unlimited supply of hydrogen, for example from Africa. 
[...]
We must continue to think about how we can use our resources more efficiently, which flights and which transports we can do without, without giving up our prosperity''}
(BUND, 143).
\end{quote}
But these more critical voices are very much a minority. In summary, we find that actors generally agree that hydrogen is relevant for climate change mitigation, but not necessarily how it compares to other mitigation options.

Economic considerations constitute another key topic. Policymakers across all parties and ministries, and incumbents from many different sectors frequently highlight the economic opportunities associated with establishing a hydrogen economy.
``\textit{We want to become the hydrogen republic of Germany. [...] And we want Germany to become the world market leader in the production and use of green hydrogen obtained from renewable energies}'' (BMBF, 123).
While policymakers stress the goal of maintaining, or even strengthening Germany’s role as a global technology exporter, several industry actors, including both incumbents and newcomers, see business opportunities in the development and export of hydrogen technology, or hydrogen-based products. However, researchers and green policymakers expect that hydrogen will remain scarce and expensive energy carrier in the future. 
``\textit{Hydrogen is the caviar among energy carriers and is too expensive and valuable to be used everywhere}'' (Green Party, 117).
In contrast, some incumbents argue either that absolute generation costs for hydrogen will further drop alongside costs of renewables, or that hydrogen will become relatively cheap if carbon prices increase. 
``\textit{The production of renewable energy, especially wind energy on the high seas, is becoming cheaper and cheaper, and at the same time the high CO2 prices, despite the Corona crisis, make coal in particular unprofitable}''
(EU commission, 118).

Technical aspects are the third key topic. Researchers and industry actors argue that hydrogen would facilitate the integration of variable renewable electricity by stabilizing and balancing the electricity grid. Incumbents from the gas, heat, and electricity sector mention the benefits of using the existing gas infrastructure to transport and store hydrogen.\footnote{The storyline encompasses both the German gas transmission grid and the distribution network, and comprises the options of blending natural gas with hydrogen, or completely converting existing natural gas pipelines to hydrogen. Conflicting views on blending natural gas with hydrogen, and using hydrogen for heating are separately discussed in Section \ref{sec:res_application}.}
``\textit{The gas customers of today are the hydrogen customers of tomorrow}'' (FNB Gas, 147).
While these storylines are largely uncontested, especially researchers and NGOs frequently emphasize the low energy efficiency of green hydrogen production and promote electrification as more efficient.\footnote{Most codes assigned to actors from the transport sector were from the CEO of the car manufacturer Volkswagen, a strong advocate for electric mobility in contrast to hydrogen based fuel cells.}

\subsection{Emerging conflicts: hydrogen use, production and imports}\label{sec:res_conflicts}

We find three emerging lines of conflict on where to use hydrogen, the production method, and imports. First, actors disagree whether hydrogen should be used widely or only for applications in DDIs with little alternatives. Second, actors disagree whether hydrogen should only be produced from renewable energies, or whether blue hydrogen is a temporary option. Finally, actors have conflicting views regarding whether hydrogen imports are rather problematic, or beneficial.

\begin{figure}[!htb]
\caption{Discourse network of key conflicts in the German hydrogen debate}\label{fig:disc_net}
\includegraphics[width=\textwidth]{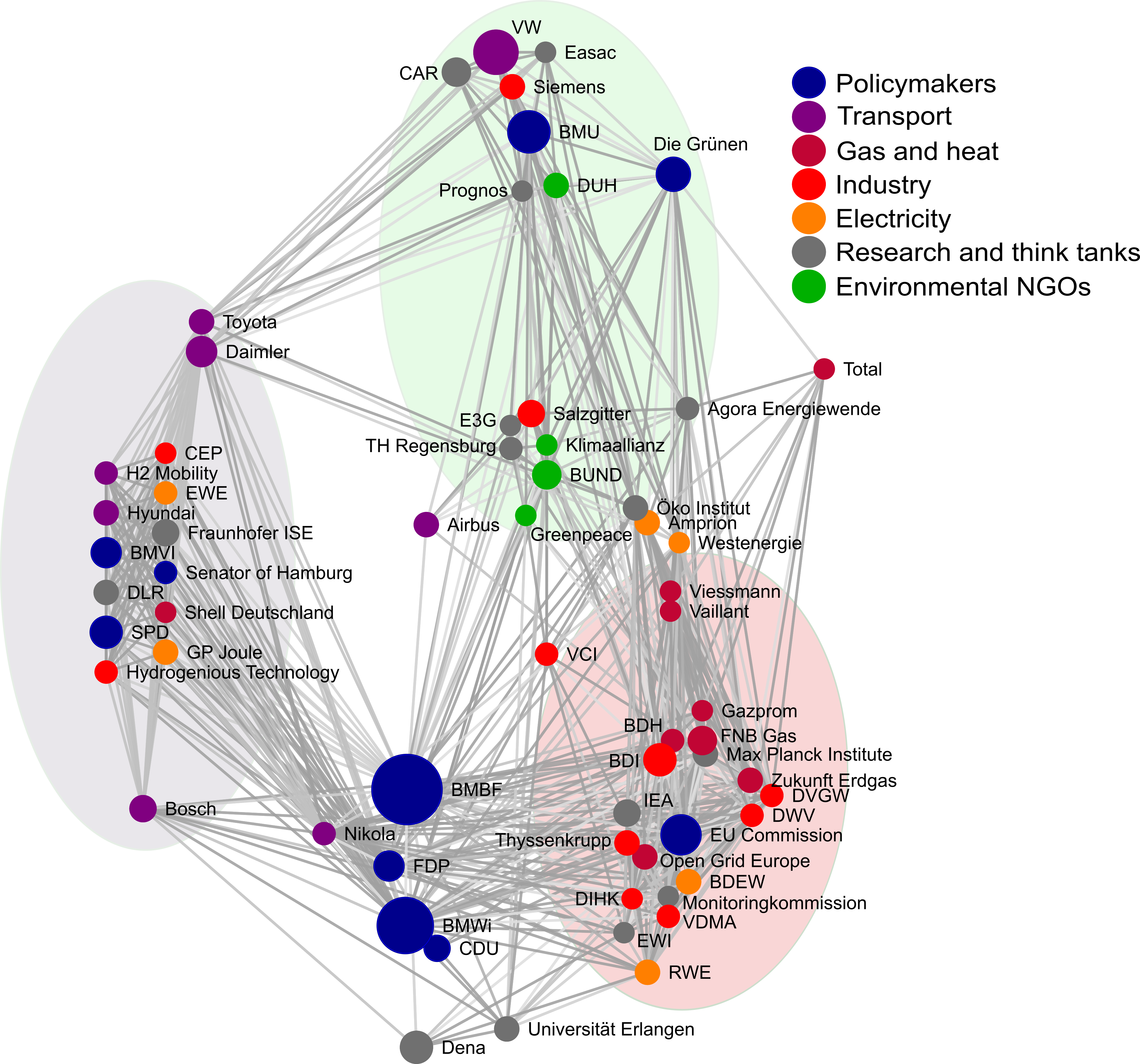}
\caption*{\footnotesize
The figure shows an actor congruence network for the three conflicts on the use, production method, and imports of hydrogen. Nodes represent actors, edges represent shared storylines. The node-size correlates with the number of newspaper articles with a coded passage by the actor. The figure only includes actors with at least three coded passages in the complete dataset. The grey shading of edges is darker for a higher normalized number of shared storylines. The node colors correspond to the actor groups, and the manually added shaded areas highlight visually identified agglomerations to facilitate the interpretation of the graph. Individual overlapping nodes were manually disentangled (especially in the grey shaded area).}
\end{figure}

\begin{figure}[!htb]
\caption{Specific lines of conflict}\label{fig:lines_of_conflict}
\begin{subfigure}{0.30\textwidth}
\includegraphics[width=\textwidth]{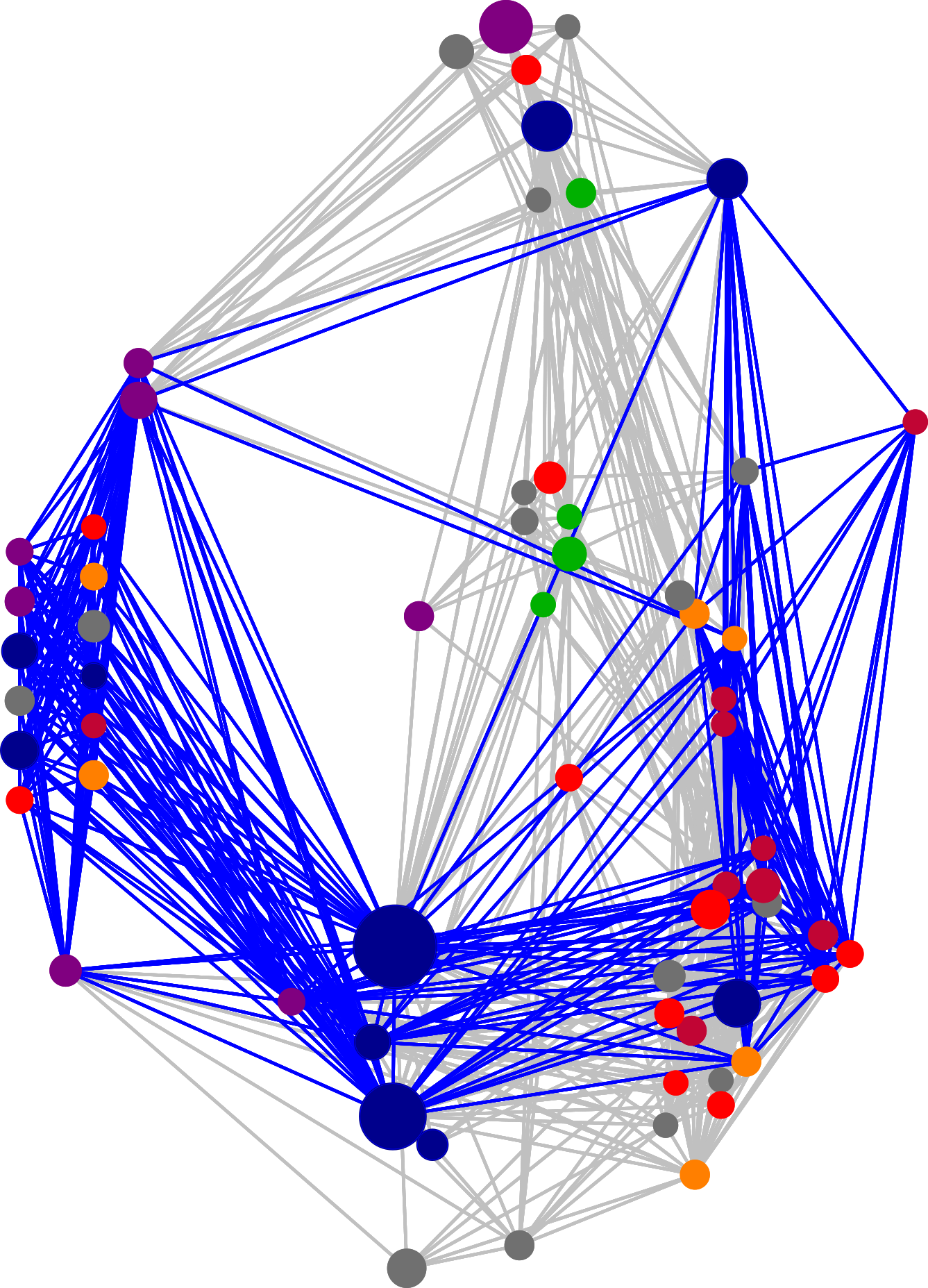}
\caption{Use: Wide use}\label{fig:actors_story_color}
\end{subfigure}
\begin{subfigure}{0.30\textwidth}
\includegraphics[width=\textwidth]{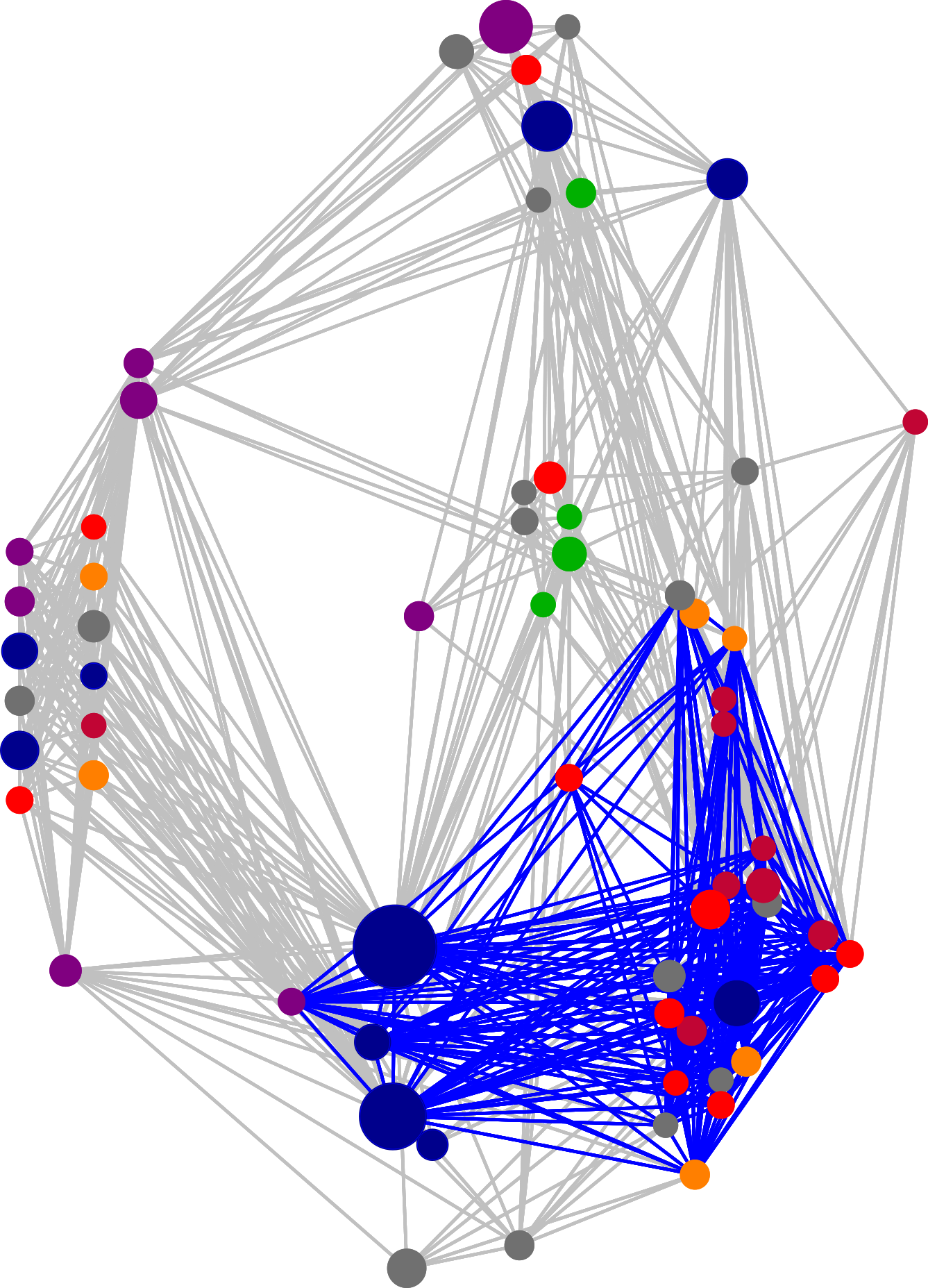}
\caption{Production method: Non-green necessary}\label{fig:actors_story_color}
\end{subfigure}
\begin{subfigure}{0.30\textwidth}
\includegraphics[width=\textwidth]{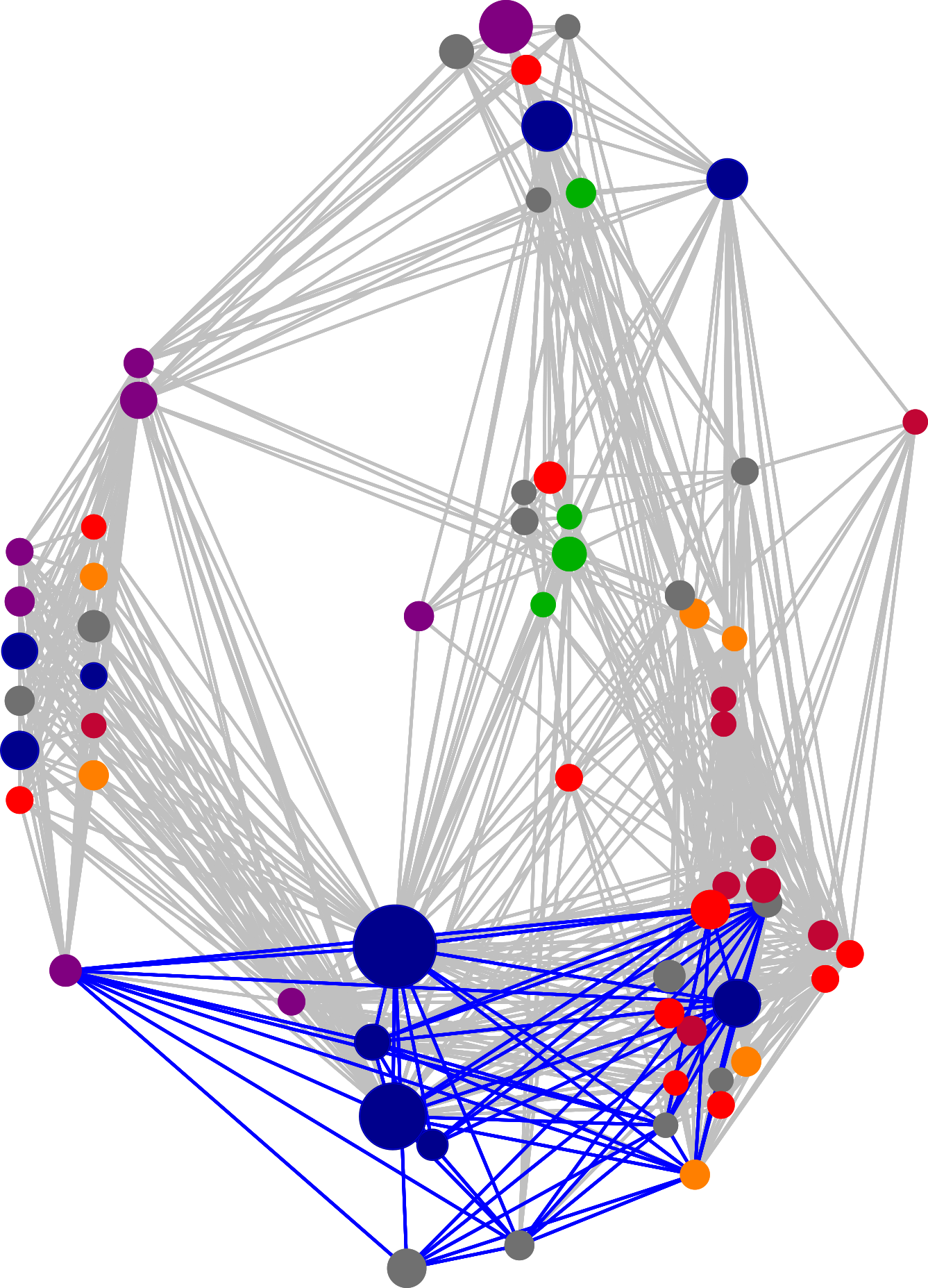}
\caption{Imports: Imports beneficial}\label{fig:actors_story_color}
\end{subfigure}
\begin{subfigure}{0.30\textwidth}
\includegraphics[width=\textwidth]{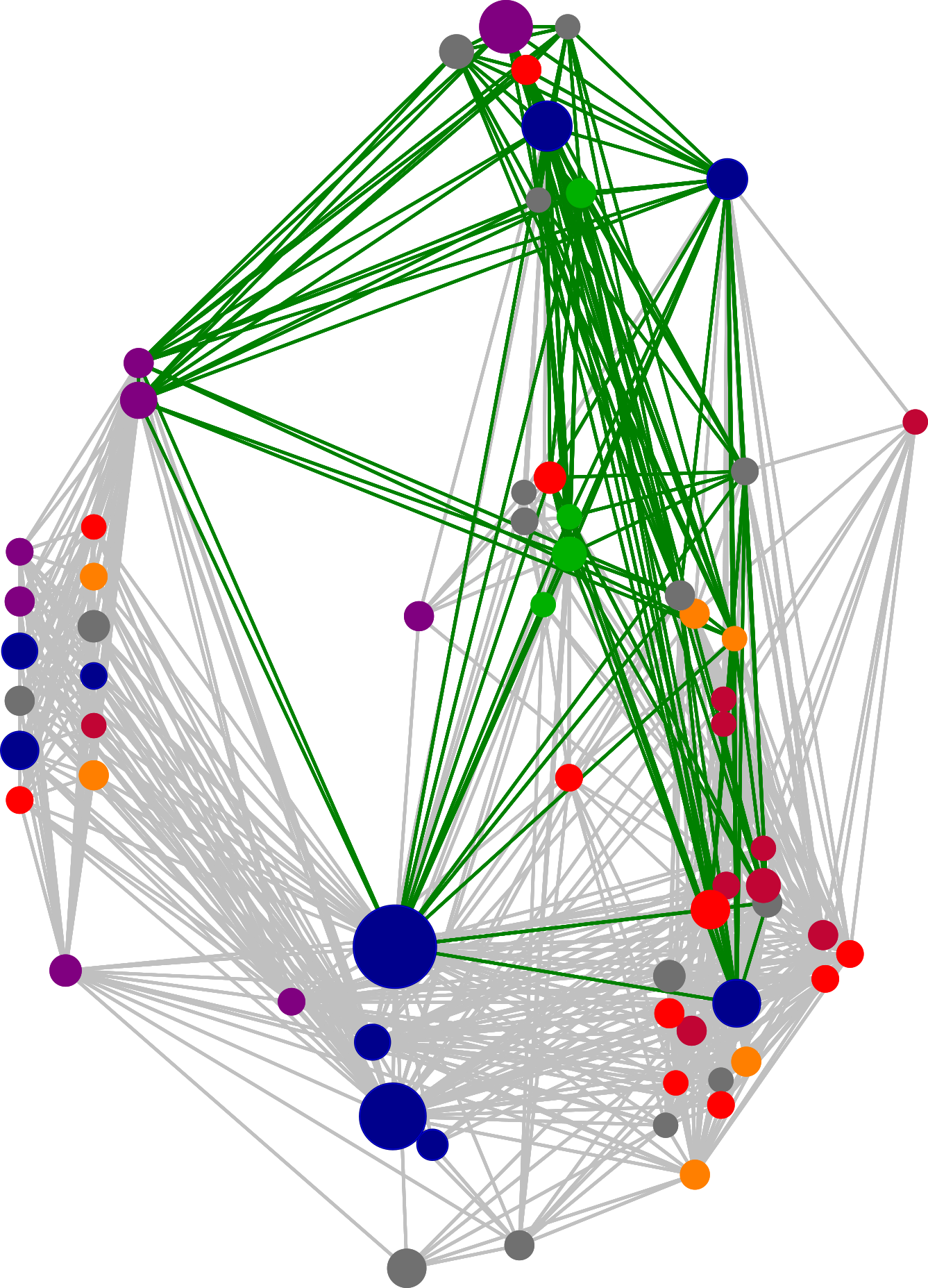}
\caption{Use: Restricted use}\label{fig:actors_story_color}
\end{subfigure}\hspace*{\fill}
\begin{subfigure}{0.30\textwidth}
\includegraphics[width=\textwidth]{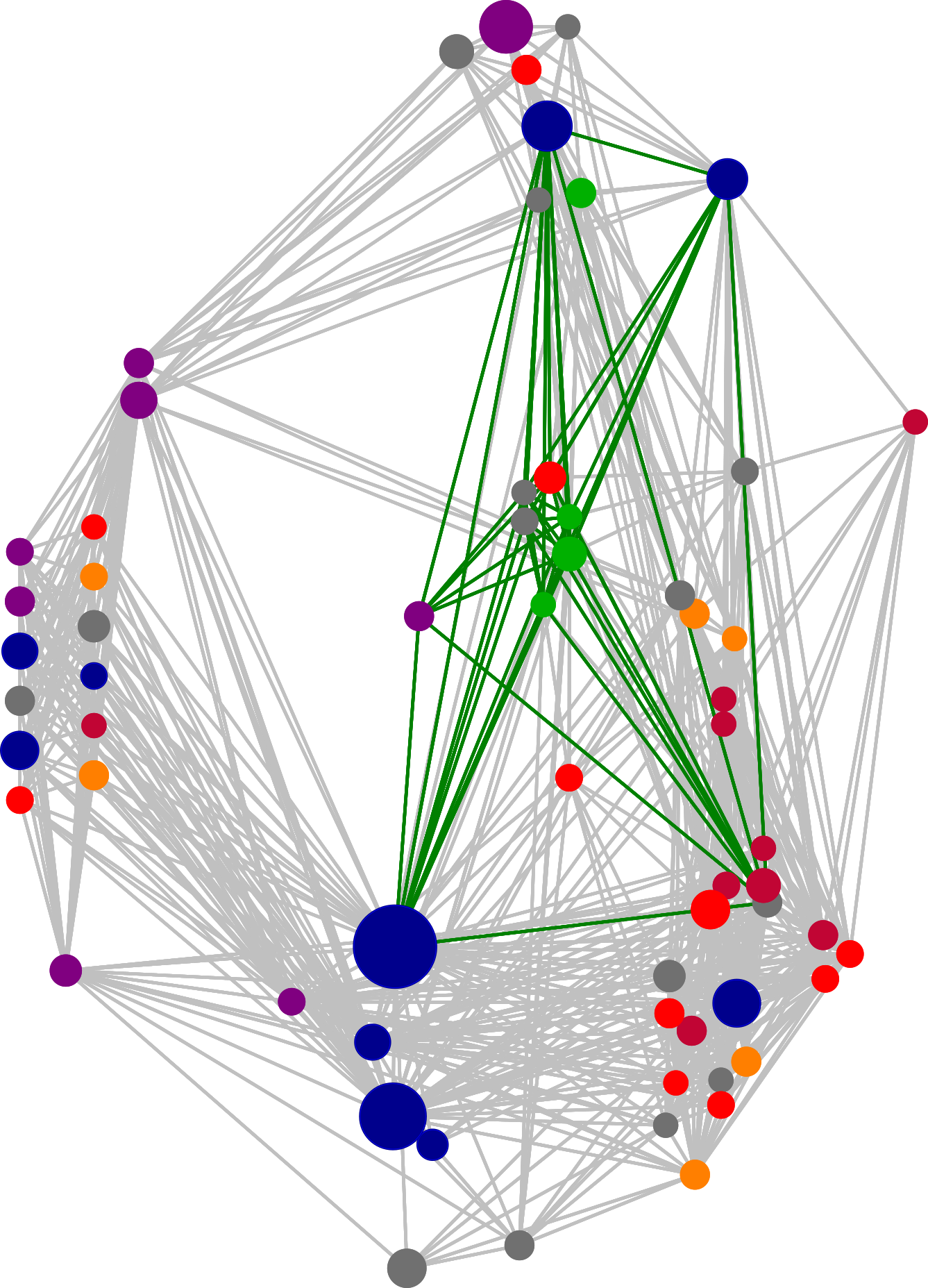}
\caption{Production method: Exclusively green}\label{fig:actors_story_color}
\end{subfigure}\hspace*{\fill}
\begin{subfigure}{0.30\textwidth}
\includegraphics[width=\textwidth]{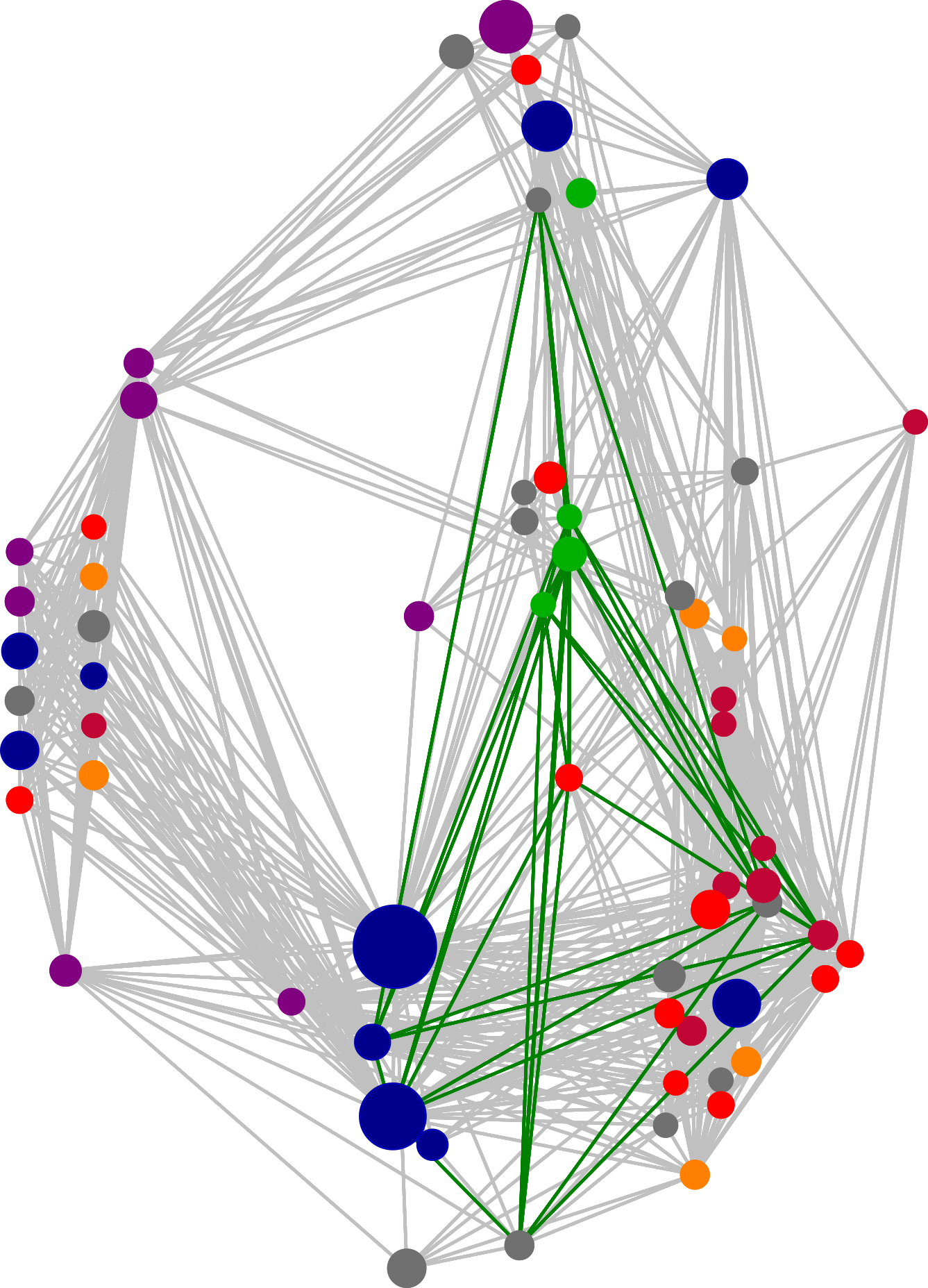}
\caption{Imports: Imports concerning}\label{fig:actors_story_color}
\end{subfigure}
\caption*{\footnotesize
The figures highlight storylines addressing the three covered conflicts that are shared by actors. The networks otherwise equals the previously shown actor congruence network (see \autoref{fig:disc_net}). The covered conflicts comprise the use (panel a and d), production method (b, e), and imports (c, f) of hydrogen (see \autoref{tab:storyline_conflicts} for more information). A panel name indicates the position within each conflict. Edges are highlighted if both actors share at least one storyline that supports the respective conflict position. Enthusiastic storylines are highlighted in blue, and skeptical storylines in green.}
\end{figure}

We employ DNA to explore the actor positions around these conflicts in more detail (see \autoref{fig:disc_net}).\footnote{The actor congruence network only includes storylines on related to the use, production,
and imports as listed in \autoref{tab:storyline_conflicts}.}
In the actor congruence network, actors are generally well connected. For a better overview, we manually added background shades to three areas with commonalities in the actor positions. Each area includes actors across groups, but some actor groups are more dominant than others: gas, heat and industry incumbents in the red area, transport sector actors in the grey and e-NGOs in the green.\footnote{The green area comprises NGOs, several research institutes and think tanks, policymakers from the German green party and the environmental ministry, a leading electric vehicle manufacturer (VW), and two leading industrial companies (Salzgitter and Siemens). The red area is dominated by incumbents from the gas, oil, power and heat sector. It also comprises the EU commission, researchers and think tanks, and actors from other industrial sectors.
The grey area is dominated by actors from the automobile sector, several policymakers, think tanks and research institutes, and hydrogen interest organizations. Some actors, especially two dominant ministries and two parties, are between the shaded areas.}
The following sections discuss the three lines of conflict in detail. 
\autoref{fig:lines_of_conflict} (a-f) highlights all edges that include at least one storyline in support of the respective conflict position.

\subsubsection{Use}\label{sec:res_application}

As a general pattern, most industry actors promote using hydrogen in their own sector, while only some argue that using hydrogen should be restricted. We classify storylines that envision using hydrogen for cars, blending it into the natural gas grid, heating, and a wide application in general, as ‘wide use’, and those that oppose one of these uses (cars, blending into gas grid, heating), or want to use hydrogen only
for DDIs, as ‘restricted use’. The specific arguments for using hydrogen differ by sector, but many outline potential benefits for the climate:
``\textit{In no other area than the heating sector would such large CO2 savings be possible so quickly and pragmatically}'' (Viessmann, 144).
Blending hydrogen into the natural gas grid ``\textit{[...] would make natural gas greener. All the investments have already been made, the pipelines work, the manufacturing plants exist}'' (Total, 179).
``\textit{We need e-fuels and hydrogen from sustainable energy sources in order to achieve the climate targets for the millions of cars in our fleet}'' (VDA, 156).
Proponents of a wide use mostly comprise actors in the red and grey areas: those in the grey area mostly emphasize using hydrogen for cars, while those in the red area propose heating, and blending hydrogen into the natural gas grid.\footnote{The Green party is an outlier due to one quote from 2019 that promotes blending hydrogen into the natural gas grid. Their position has changed since, which is why it also features prominently among those that want a more restricted use. We find contradicting storylines also for other actors due to different individuals being aggregated, a temporal change in the actor’s position, or simply inconsistent communication.}

Those favoring a restricted use argue that hydrogen should primarily be used in sectors with few alternatives for decarbonization. Otherwise hydrogen might be missing where it is needed most.
\begin{quote}
``\textit{The [steel] industry has no alternative if it wants to become climate-neutral. 
[...] The cement industry and the chemical industry are also dependent on hydrogen. In air and sea transport, hydrogen is also the central building block on the way to climate neutrality}'' (BMU, 80).
\end{quote}
Some actors also explicitly advise against using hydrogen in specific applications.
``\textit{Green hydrogen has no place in cars and heating systems}'' (DUH, 104).
Storylines supporting a restricted use are prominently located in the green area, with some exceptions in the red area: For example, the EU Commission explicitly prioritizes hydrogen for industrial uses. Toyota and Daimler (grey area) are also noteworthy, as both predominantly promote hydrogen for trucks or buses, not for private cars.

\subsubsection{Production method}\label{sec:res_prod_method}

This conflict centers around two major arguments: one is about only producing hydrogen from renewable energies, and the other is about making large quantities of hydrogen available quickly by also using fossil fuels. Many actors argue that blue hydrogen is required for a transition period, or simply state that various forms of production should be considered. ``\textit{Hydrogen has many colors, and we should use all. [...] We should use blue or turquoise hydrogen for a transition period}'' (RWE, 115).\footnote{One storyline of (a) states that ``blue hydrogen was required for a transition period or building up the market, but that green hydrogen would be the ultimate goal''. This storyline is practically in between both positions. However, we distinguish between actors that agree to building infrastructure for non-green hydrogen as well, and those who disagree.}
A variation of this argument is that only green hydrogen would not be possible due to limited availability or high costs. A minority even considers grey hydrogen as a temporary solution. Those in favor of blue hydrogen include almost all actors in the gas and heat sector, several policymakers, and also a leading environmental research institute: ``\textit{[I]n the next decade [blue hydrogen will] be the only source that is justifiable in terms of cost for high-volume applications and for the market ramp-up in those areas in which hydrogen is of high strategic importance}'' (Öko-Institut, 171).
Also the EU Commission suggests using blue hydrogen.

Opponents of this position argue that only green hydrogen is actually carbon free. Some also mention that carbon capture and storage (CCS) is a contested technology: ``\textit{Why should we use blue hydrogen in the future, if the climate footprint is bad and the costs for generation high? 
[...]
It instantly brings a debate about CCS [...], which is predominantly rejected in Germany}'' (BMU, 80). 
A few also highlight that using blue hydrogen would be a lifeline for the fossil fuels: ``\textit{Blue hydrogen is of fossil origin and perpetuates the fossil industry instead of transforming it}'' (TU Regensburg, 112).
These storylines are predominantly assigned to actors in the green area including the environmental ministry, and to the ministry for education and research. A leading steel producer (Salzgitter) even wants to produce its green hydrogen for its steel production with own renewable power plants, thus taking a dual role of using and producing hydrogen.

\subsubsection{Imports}\label{sec:res_imports}

Actors arguing that hydrogen imports are beneficial explain that hydrogen generated abroad would profit from better solar conditions and is hence potentially cheaper, or that exporting countries would benefit.
\begin{quote}
``\textit{With
green hydrogen, the geographical advantages in renewable energies could become a development engine for the societies there [Africa]. 
[...]
In this way, we not only create the basis for German technology exports, but also ensure a climate-friendly energy supply}'' (BMBF, 61).
\end{quote}
Benefits of hydrogen imports are mentioned by a leading industry association (BDI), the EU Commission, several research institutes in the red area, as well as by several policymakers, and a large car industry supplier (Bosch).

Those concerned of hydrogen imports warn of new import dependencies, and of potential environmental and social risks for exporters. 
``\textit{Importing hydrogen from countries of the Global South without adequate consideration of the ecological and social situation in the country of production risks being perceived as a mechanism of exploitation or a new form of colonialism}'' (Brot für die Welt, 173). ``\textit{Many of the countries that basically come into consideration still have to develop themselves first. [...] They would not export the green hydrogen, but use it for their own economic development}'' (Zukunft Erdgas, 79).
Concerns about imports are mostly raised by policymakers, NGOs and research institutes. Risks for exporters are highlighted by NGOs and members of the German liberal party (FPD), while some incumbents (Zukunft Erdgas and VCI) warn of import dependency.

\subsection{Summary along actor groups}\label{sec:res_conflict_comparison}

The public debate about hydrogen in Germany is characterized by a broad agreement that hydrogen is necessary for achieving net-zero targets. The topics of climate change mitigation and creating economic opportunities receive most attention (123 and 87 coded passages). Technical aspects and storylines related to costs of hydrogen are less prevalent (71 and 27 coded passages). Overall, we find that each actor group supports hydrogen for overlapping, but also partially different reasons. The ambiguity around what a ‘hydrogen economy’ actually implies might explain some of the observed enthusiasm.

Most industry actors are very much in favor of hydrogen and emphasize the prospect of cheap hydrogen, or the opportunity to utilize the existing gas infrastructure. For these actors, policy support for hydrogen projects (in their respective sector) is particularly important, as they are confronted with large upcoming investments, but often bound to existing assets and business models. The broad support by industry incumbents is particularly interesting because they vary substantially in terms of sectors, technologies and strategies (see Section 5.1).

NGOs are more skeptical with their support of hydrogen. NGOs highlight the limitations of hydrogen (costs, scarcity, efficiency) and other, potentially neglected decarbonization options. Along these lines, NGOs, together with the ministry for the environment and the Green Party, favor a restricted use of hydrogen, exclusively support green hydrogen. They also emphasize risks for potential exporters, a topic with little salience beyond. NGOs may want to remind policymakers and the public of looming problems and potential caveats.

Policymakers mostly emphasize the economic opportunities associated with hydrogen, although a few also mention that it might remain scarce and expensive. For policymakers, hydrogen is a very favorable topic that creates many winners, allows for prestigious projects, and does not yet require difficult decisions or priority setting at this early stage. Regarding the use and production method, policymakers are divided, depending on their party affiliation. They address both potential risks and benefits of hydrogen imports. Policymakers thus overall seek to create a positive image around the prospects of a hydrogen economy.

Researchers and think tanks engage with all topics. They all highlight the necessity of developing hydrogen technologies, while being explicit about the limitations and caveats. Regarding the three conflicts, their positions vary. There are those that very much concentrate on the techno-economic characteristics. Apart from some enthusiasts, there are many cautious voices that highlight the technical limitations of different options.

\section{Discussion}\label{sec:discussion}

\begin{figure}[!htb]
\caption{Applied conceptual framework}\label{fig:concept_frame_appl}
\includegraphics[width=\textwidth]{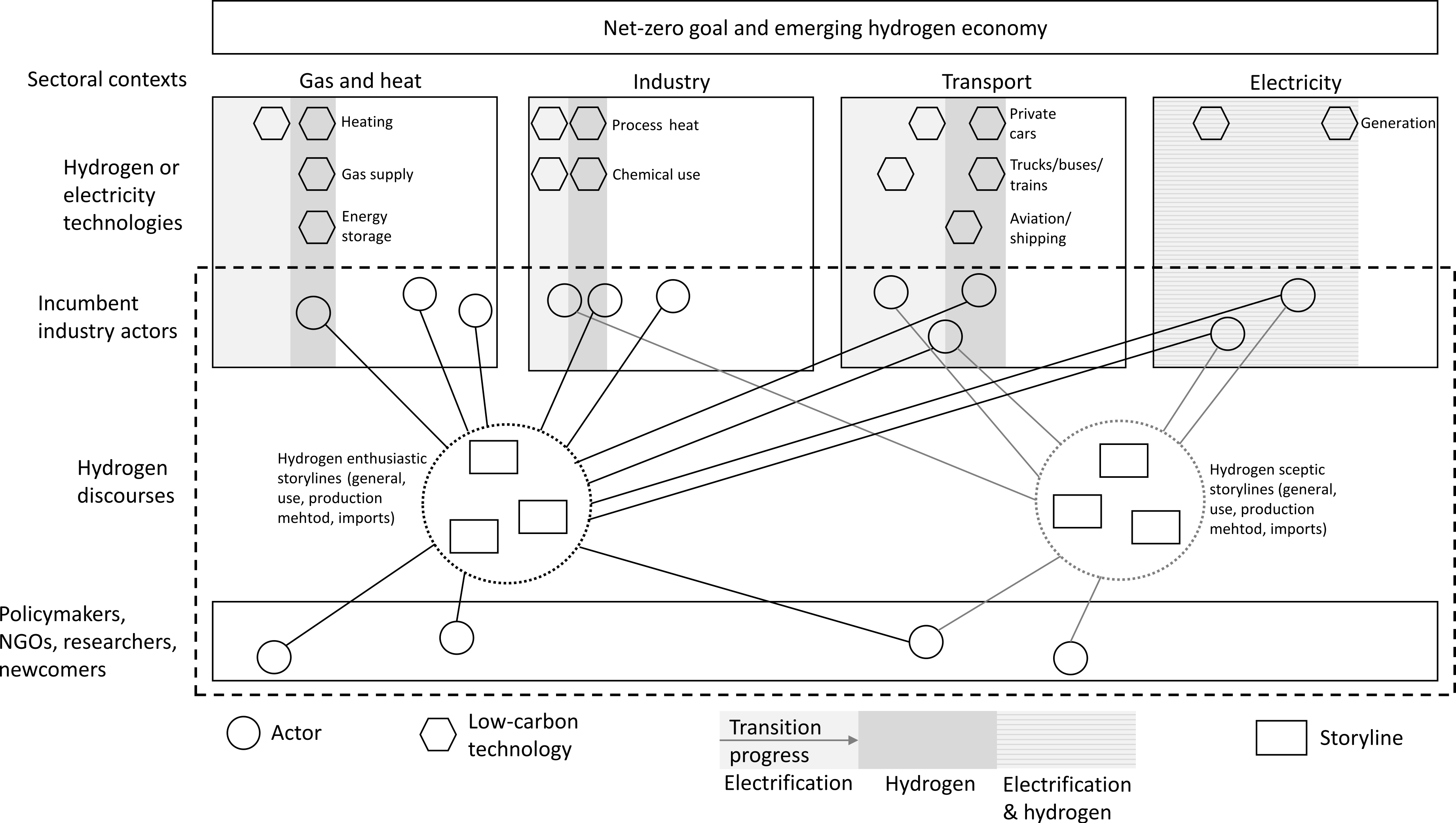}
\caption*{\footnotesize
The applied conceptual framework positions actors in the sectors gas and heat, industry, transport, and electricity in a stylized manner in the discussion. The sectors are at different transition stages towards using technologies based on electricity or hydrogen. The depicted technologies only build a selection to illustrate the overall picture. The different low-carbon technologies are competing, except for the electricity sector, where both complement each other. In line with our empirical analysis, we only focus on the discussion around hydrogen.}
\end{figure}

Our findings show that many actors, including incumbents from different sectors and industries support the idea of establishing a hydrogen economy in Germany. The enthusiasm by incumbents is particularly interesting for the transitions literature, given that they have been observed playing different roles (see Section \ref{sec_actors}). To better understand their motivations, we seek explanations at three levels: 
sectoral contexts, sector-specific technology features (including the availability of alternatives to hydrogen), and actor interests (see Section \ref{sec:concept_framework} and \autoref{fig:concept_frame_appl}). In our case, we look at the intersection of four different sectors. In each sector, there are different technological options for decarbonization in different stages of development. We discuss electrification and hydrogen in the following (see \autoref{tab:sectoral_transitions}). For the sake of simplicity, we assume just two general positions: enthusiastic vs. skeptical.

\begin{landscape}

{
\def\sym#1{\ifmmode^{#1}\else\(^{#1}\)\fi}


\captionof{table}{Transitions in each sector} \label{tab:sectoral_transitions}

\footnotesize
\onehalfspacing

\renewcommand{\arraystretch}{1}

\setlength{\tabcolsep}{3pt}

\begin{xtabular}{p{3cm}p{10cm}p{6cm}p{6cm}}

\hline
\rule{0pt}{3ex} \textbf{Sector} & \textbf{Sectoral context}   & \textbf{Sector-specific technologies}  & \textbf{Actor interests} \\
\hline
\textbf{Gas and heat}    & 
\colorbox{coolgrey}{Hydrogen:} Transition at an early stage. Increasing public and political pressure to decarbonize the building sector and phasing-out natural gas. Policies are under development, upcoming regional feed-in of hydrogen to gas grids. Potential for local and national power storage. Energy transport via pipeline to avoid contested deployment electricity deployment of transmission power grid. 
\colorbox{battleshipgrey}{Electricity:} Transition at medium stage. Low-carbon heating supported by public funding schemes.         & 
\colorbox{coolgrey}{Hydrogen:} Heating via blending hydrogen into existing natural gas grid, or redesigning natural gas infrastructure. Large potential energy storage and transmission capacities in existing gas infrastructure. Infrastructure potentially needed for gas supply of combined heat and power generation (see Sector Electricity).  
\colorbox{battleshipgrey}{Electricity:} Heat-pumps are economically competitive and increasingly deployed.         & 
\colorbox{coolgrey}{Hydrogen:} Strong support by incumbents, as hydrogen only option to sustain current business models, with potential of blue hydrogen to save natural gas assets.                 
\colorbox{battleshipgrey}{Electricity:} Not supported by gas incumbents due to conflict with business model. Supported by manufacturers of heat-pumps though.        \\
\textbf{Industry}                                     & 
\colorbox{coolgrey}{Hydrogen} and \colorbox{battleshipgrey}{Electricity:} Transitions at early stages. Difficult-to-decarbonize industries, such as steel, chemical and cement industry, of increasing political focus. Some industries under strong international competition. Hydrogen considered primary solution, large public transition funding envisaged e.g. via carbon-contracts for difference.
                                                                                         & 
\colorbox{coolgrey}{Hydrogen} and \colorbox{battleshipgrey}{Electricity} based technologies still under development. Hydrogen for production processes requiring high temperatures and chemical properties of energy carrier, electricity only for high temperatures.                            & 
\colorbox{coolgrey}{Hydrogen} and \colorbox{battleshipgrey}{Electricity:} Incumbent actors generally support using hydrogen or electricity, depending on industrial process and political support. No newcomers involved.
                    \\
\textbf{Transport}                        & 
\colorbox{coolgrey}{Hydrogen} and \colorbox{battleshipgrey}{Electricity:} Especially road transport under increasing decarbonization pressure due to national and EU regulation. Often both hydrogen and electricity technically possible. \colorbox{coolgrey}{Hydrogen:} Transition at early-medium stage. Fuel cell cars scarcely sold. First commercial projects for heavy transport and trains. First pilot projects for shipping and aviation.   
\colorbox{battleshipgrey}{Electricity:} Transition at medium stage. Battery electric vehicles with increasing market shares, increasing charging infrastructure. First commercial projects for heavy transport and trains. No option for aviation, first pilot projects for electric ships. & 
\colorbox{coolgrey}{Hydrogen:} Fuel cell cars fully developed but expensive. Fuel cell trucks, buses and trains developed. Technology for aviation and shipping under development.
\colorbox{battleshipgrey}{Electricity:} Battery electric vehicles fully developed. Electric trucks, buses and trains developed. No option for aviation, ships for short distance under development.    & 
\colorbox{coolgrey}{Hydrogen:} Support ranges from weak to strong, depending on application, and individual decisions by companies on specific technology. Support generally weaker for smaller and lighter vehicles. Incumbents challenged by newcomers (e.g. Nikola).
\colorbox{battleshipgrey}{Electricity:} Support inverse to hydrogen, generally stronger for smaller and lighter vehicles. Incumbents challenged by newcomers (e.g. Tesla). \\
\textbf{Electricity}                                  & 
Special context: Energy carriers complementary, electricity primary option. 
\colorbox{battleshipgrey}{Electricity:} Transition at medium-late stage: Share of renewable energies increases, although deployment slowed down by local resistance and lengthy project implementation. Coal
phase-out decided, natural gas increasingly less of considered transition option.                                                            
\colorbox{coolgrey}{Hydrogen:} Thus far, no commercial power production based on hydrogen, only expected for energy systems with high shares of renewables.
     & 
\colorbox{battleshipgrey}{Electricity:} Renewable energies generate power, most importantly PV and wind. 
\colorbox{coolgrey}{Hydrogen:} Natural gas power plants available for power generation to balance electricity grid, but need adjustment for hydrogen. & 
\colorbox{battleshipgrey}{Electricity:} Support by most incumbents as portfolio now includes renewable technologies. Demand adjusted regulations for faster deployment of renewables.  
\colorbox{coolgrey}{Hydrogen:} Support by some incumbents as new business field through green hydrogen production.         \\
\hline
\end{xtabular}
\caption*{\footnotesize
The table describes 
the \textit{Sectoral context}, 
\textit{Sector-specific technologies}, 
and \textit{Actor interests} in the sectors 
Gas and heat,
Industry,
Transport, and 
Electricity.}
}
\end{landscape}

We now elaborate the actor positions in each sector (see Section \ref{sec:sectoral_transitions}), outline implications for transitions research (see Section \ref{sec:discussion_categories}), and finally discuss limitations to our approach (see Section \ref{sec:limitations}).

\subsection{Explaining actor positions for each sector}\label{sec:sectoral_transitions}

Gas and heat sector incumbents are among the most vivid supporters of a large hydrogen economy (see \autoref{tab:sectoral_transitions}). Decarbonization of heating has not progressed well but there is an ongoing transition (medium stage) toward electrification of heating (e.g. with heat-pumps). In the past, incumbent actors have promoted gas as an alternative to oil, now they argue for ‘green gas’ instead of electrification \citep{lowes_heating_2020}. Hydrogen fits well into this strategy. It would allow gas suppliers to continue managing the grid infrastructure and boiler manufacturers to continue to sell boilers. Gas producers could even use their natural gas sources to produce blue hydrogen. Overall, actors in the gas and heat sector try to create an image of hydrogen as a pragmatic, fast, and low effort solution to decarbonize the heating sector. This strategy downplays cheap, efficient and available technological alternatives (heat pumps, insulation) and helps incumbents to keep control of their assets and business models.\footnote{The above strategies and the role of natural gas suppliers might fundamentally change in response to the Russian invasion of the Ukraine and the ensuing issues around gas prices and security of supply.}

Incumbents of Germany’s energy intensive industries (e.g. steel, chemical and cement industry) also want to decarbonize their businesses with hydrogen. In most industries, decarbonization (with hydrogen or electricity) is still at early stages and incumbent actors face great pressures. Hydrogen is frequently highlighted as a key, or even the only solution, especially for processes such as steel making or ammonia or plastic production that require hydrogen molecules. Processes that require high temperatures may also be decarbonized via electricity \citep{madeddu_co_2020}. Incumbent actors therefore support the idea of establishing a hydrogen economy, suggest a fast ramp-up and expect political support \citep{bmwk_habeck_2022}. An important industry association (BDI) supports using blue hydrogen for an intermediate period and suggests
not to use hydrogen for heating because of limited supply. Viewpoints on imports vary. Some actors expect benefits due to lower costs (BDI), while others fear import dependencies (VCI). In the public debate, hydrogen is presented as the only solution for a successful transition of the German industry, potentially even creating a competitive advantage.

Incumbents in the transport sector are divided between supporting hydrogen or electricity. Transition stages differ depending on the mode of transport and energy carrier. Hydrogen based solutions are at very early stages for ships or airplanes and more developed for cars, trucks and trains. The transition to electric vehicles is progressing rapidly for cars and more slowly in the case of trains or trucks. There are also first electric airplanes. The entire transport is under increasing political pressure and incumbents are also challenged by newcomers, such as Tesla or Nikola. In the public debate, incumbent actors mostly discuss specific technological aspects and emphasize the potential of hydrogen for specific applications. Some German car manufacturers present hydrogen as a technologically superior (or even the only) long-term solution with advantages over electricity, while others point to the rapid progress in electric vehicles. Engagement in more general conflicts about the production method or imports are scant.

Actors from the electricity sector generally support the deployment of hydrogen. Decarbonization on the basis of renewable energies is comparatively advanced and fossil fuels are envisaged to be phased-out eventually. In the electricity sector, in contrast to the other sectors, electricity and hydrogen do not compete but may complement each other. Hydrogen can be used to store and balance the supply of fluctuating renewables, while renewables are needed to produce green hydrogen. Most incumbents are in favor of using green hydrogen because this provides new business opportunities. Some companies also promote blue hydrogen for a transition period, and using hydrogen for heating. The image created around hydrogen is that of an inevitable technological component facilitating the transition in electricity and providing new business opportunities.

\subsection{Implications for transitions research and policy}\label{sec:discussion_categories}

Our analysis has shown that hydrogen and the net-zero energy transition bring a new degree of complexity to transition studies. Here we discuss broader implications of multi-system interaction, multi-technology interaction, and strategies of (incumbent) actors. 
First, we need to embrace multi-system interactions and multiple transitions unfolding in parallel across different sectors \citep{rosenbloom_engaging_2020, kanger_research_2021}. Each sector has its own specificities in terms of sustainability challenges, actors and interests, regime structures and approaches for decarbonization. Also, we have seen that sectors are in different transition stages and that developments in more advanced sectors such as electricity affect transitions in less advanced sectors. 
For the transition to net-zero it is important that transition pathways in different sectors complement each other to generate cumulative effects towards decarbonization. However, our study has shown that actors have conflicting views about where to use hydrogen or how much hydrogen to import. This means that frictions can occur as actors compete for scarce resources or political support. Such conflicts will require careful priority setting through policy, e.g. by defining applications and sectors that shall get hydrogen first.

Second, we need to consider multi-technology interactions, e.g. between technologies at different stages of maturity \citep{papachristos_system_2013, andersen_multi-technology_2020}. 
In our case, hydrogen and electricity can both be the basis for many low-carbon technologies, which means hydrogen and electricity compete but may also complement each other. Depending on the sector and application, actors will also perceive different technologies as more or less disruptive. While we focused primarily on hydrogen, future studies might want to analyze several innovations at the same time to explore these interactions further.

Third, new conceptual frameworks need to accommodate a broad variety of actors with different strategic interests. We observe that incumbents’ position towards hydrogen seems to be affected by their sectoral background, the performance and level of disruption of new sector-specific technologies, and prior strategic decisions. We observe support for hydrogen in sectors where i) it is (currently) the only low-carbon option, ii) it complements existing technologies, or iii) it is compatible with the status quo business model of incumbents. Independent of the sector, it also receives support if it constitutes a new business opportunity. Reactions are more diverse in sectors where some incumbents work with electricity while others prioritize hydrogen. Future frameworks could more systematically categorize such relations.

Our analysis revealed broad support of hydrogen by a wide range of incumbent actors, which might be counterintuitive. While hydrogen is a case in which incumbents drive innovation \citep{berggren_transition_2015, turnheim_forever_2020}, we also find elements of ‘politics of delay,’ e.g. by actors from the gas and heat sector \citep{lamb_discourses_2020}. Such settings of a complex interplay of various strategic interests require careful policymaking to not risk capture by powerful incumbents.

Finally, our case also points to the potential emergence of meta-rules and transition patterns that span across multiple systems \citep{kanger_deep_2019}. One meta-rule that is already prominent in the net-zero energy transition is low-carbon electrification, with more and more applications being converted from fossil fuels to electricity. In the future, hydrogen may assume a similar role: like electricity, it might become a widespread basis for decarbonization. Future studies will have to explore the benefits (e.g., complementary use of infrastructures) but also risks (e.g., lock-ins) of these developments.

\subsection{Limitations}\label{sec:limitations}

Our study is subject to the following limitations. First, what we learned for Germany does not necessarily apply elsewhere. For example, the German hydrogen strategy envisages large-scale hydrogen imports, and only supports green hydrogen. We expect different discourses in places that envision becoming large-scale exporters of green hydrogen, or countries with abundant domestic natural gas resources that could benefit from producing blue hydrogen. 

Second, we only analyzed the debate about hydrogen, but neglect interrelated discourses on other net-zero options, for example low-carbon electricity, or negative emission technologies. Future studies could explore the politics around multiple technologies for decarbonization in greater detail. 

Third, as the hydrogen debate emerged rather recently, we were not able to study how actor positions and arguments changed over time.\footnote{We describe arising limitations due to limited data in Section \ref{sec:data_analysis}.}
It will be very interesting to repeat our analysis in a few years’ time.

Fourth, we only analyzed the public debate through newspapers. Other venues and data sources such as parliament debates, position papers, or expert interviews may reveal higher levels of detail as well as different actor interests in terms of audience and the aspired goal of communication. We also acknowledge that newspaper articles are subject to journalists’ pre-selection of information.

Finally, storylines aggregate distinct statements to provide a comprehensive overview, at the risk of neglecting sector-specific debates and nuances of different arguments. Future studies could analyze larger bodies of data using machine-learning based coding techniques to compare hydrogen discussions across different countries, or to consider multiple net-zero technologies.

\section{Concluding remarks}\label{sec:concluding_remarks}

The net-zero energy transition calls for swift and economy-wide action with many sectoral transformations unfolding in parallel and a broad range of technologies and actors involved. In the case of hydrogen, we found a broad range of actors, including many incumbent firms from different sectors with diverging preferences for specific technology solutions. Interestingly, many incumbent actors support the development of a hydrogen economy, while environmental NGOs and several think tanks are less enthusiastic. We explain these findings with sector-specific constellations of low-carbon alternatives, different transition stages and firm-specific strategic decisions. The complex interplay of multiple sectors, technologies and actors creates new challenges for policy and research. Our conceptual framework helps to untangle and identify different levels (sector, technology, firm), different elements at each level (e.g. multiple sectors) and some key parameters (e.g., transition stage, availability of technology alternatives, strategic decisions) affecting the discourses in the public debate.

\section{Acknowledgements}
 
Nils Ohlendorf gratefully acknowledges funding by the German Federal Ministry of Education and Research (PEGASOS, funding code 01LA1826C). 
Meike Löhr acknowledges funding by the German Research Foundation (grant no. 316848319) in the context of the Emmy Noether-project “Regional energy transitions as a social process” (REENEA).
Jochen Markard acknowledges funding from the Norwegian Research Council (Conflicting Transition Pathways for Deep Decarbonization, Grant number 295062/E20) and from the Swiss Federal Office of Energy (SWEET program, PATHFNDR consortium).

We thank participants and discussants at the International Sustainability Transitions Conference (IST 2021),
5th International Conference on Public Policy (ICPP5),
the Sustainable Technologies Colloquium at the ETH Zürich,
and members of the Climate and Development group at the MCC Berlin for their helpful comments and suggestions.

\newpage

\addcontentsline{toc}{section}{References}
\renewcommand\refname{References}
\bibliographystyle{apalike}
\bibliography{Literature,Literature_alt}

\newpage

\appendix
\section{Appendix}\label{appendix}

\setcounter{table}{0}
\renewcommand{\thetable}{A.\arabic{table}}

\setcounter{figure}{0}
\renewcommand{\thefigure}{A.\arabic{figure}}

\subsection{Search query and newspaper articles}\label{appendix:search_query}

We obtain the newspaper articles for our analysis by conducting a systematic search across different databases. The query includes articles with hydrogen mentioned at least once at the beginning, and four times in the main text. We only include articles that exceed 300 words. The precise time period ranges between the 01.01.2016 and 31.12.2020. Unrelated articles are removed from the search via a number of characteristic keywords. The following search query was used for the database Lexis, from where we obtain articles for \textit{Die Welt}, \textit{Frankfurter Allgemeine Zeitung} and \textit{TAZ}:

hlead(Wasserstoff*) AND ATLEAST4(Wasserstoff*) AND Publication(Frankfurter Allgemeine Zeitung OR taz OR Die Welt) AND länge$>$300 AND date $>$ 12/12/2015 AND date $<$ 31/12/2020 AND NOT (Fusion* OR Zirkon OR Neutron* OR Helium OR Antimaterie OR myon* OR Deuterium OR Sterne OR MRT OR Wasserstoffperoxid OR Graphen OR Wasserstoffbombe OR Stempelzelle OR ``Kryo-Wasserstoff'') AND NOT Publication(Die Welt Hamburg)

The queries for WISO (\textit{Handelsblatt}) or the \textit{Süddeutsche Zeitung} archive are analogue, except for the following deviations:
WISO prevents including a ``*'' to the query element that sets a minimum number of keyword mentions, resulting in potentially less findings. 
The Süddeutsche Zeitung query only receives articles with the keyword mentioned in the article title, without the option to set a minimum requirement for keyword mentionings in the full text, leading to potentially more findings.

They searches yield 321 articles in total. 
The articles split as to newspapers as following:
\textit{FAZ} 103, 
\textit{Handelsblatt} 75, 
\textit{SZ} 61, 
\textit{Die Welt} 49 and 
\textit{taz} 33.
The final sample of 179 articles emerges after removing 
i) 3 articles from 2015,
ii) 21 duplicates or false hits,
iii) 34 articles without any content code,
iv) 57 articles that only contains codes by journalists or actors omitted from the analysis,
and, finally, v) 30 articles that exclusively contained storylines omitted from the analysis.
The number of storylines thereby reduces from almost 4000 initially coded passages, 
to 614 in the final analysis,
mostly during step iv) and v). 
The same storyline can be coded only once per article by the same actor.
\autoref{tab:newspaper_articles} lists all newspaper articles.

{
\def\sym#1{\ifmmode^{#1}\else\(^{#1}\)\fi}


\captionof{table}{Newspaper articles} \label{tab:newspaper_articles}

\footnotesize
\onehalfspacing

\renewcommand{\arraystretch}{1}

\setlength{\tabcolsep}{2pt}

\begin{xtabular}{p{0.5cm}p{2cm}p{2cm}p{11cm}}

\hline
\rule{0pt}{3ex}N & Date & Newspaper & Title \\ 
\hline
1            & 1/29/2016        &           SZ & Gefesselte Energie                                                                                                                                                                                                                                                                                                                        \\
2          & 3/9/2016           & Handelsblatt & Taktgeber für Technik                                                                                                                                                                                                                                                                                                                                  \\
3          & 4/13/2016          & Die Welt     & Nur heiße Luft; Der Verkehrsminister will Wasserstoff-Autos fördern. Doch   die Strategie ist ein Aufguss alter Pläne                                                                                                                                                                                                                                  \\
4          & 5/6/2016           & Handelsblatt & Thomas Bystry; 'Es sind nicht mehr viele Fragen offen'                                                                                                                                                                                                                                                                                                 \\
\hline
\end{xtabular}
\caption*{\footnotesize 
The table shows the identifying number,
date, 
newspaper, 
and title of each newspaper article included to the analysis.}
}


\subsection{Actors}\label{appendix:descr_actors}

\autoref{tab:actor_overview} lists all 
actors, including the number of 
coded storylines,
articles the actor appears in, 
and different storylines coded for the respective actor.
It also shows to which group each actor is assigned to.
The public debate is rather concentrated as few actors show significantly more codes than others, 
and appear in more newspaper articles.

{
\def\sym#1{\ifmmode^{#1}\else\(^{#1}\)\fi}


\captionof{table}{Actor overview} \label{tab:actor_overview}

\footnotesize
\onehalfspacing

\renewcommand{\arraystretch}{1}

\setlength{\tabcolsep}{2pt}

\begin{xtabular}{p{5cm}p{2cm}p{2cm}p{2.5cm} p{4cm}}

\hline
\rule{0pt}{3ex}Actor name & Codes & Articles & Different SLs & Group\\ 
\hline
BMBF                         & 55             & 26                & 13       & Policymakers              \\
EU Commission                & 28             & 11                & 11       & Policymakers               \\
BMWi                         & 27             & 19                & 11      & Policymakers                \\
BMU                          & 21             & 12                & 10      & Policymakers                \\
Max Planck institute                   & 20             & 3                 & 16      & Research and think tanks               \\
BUND                         & 19             & 5                 & 10          & NGOs           \\
VW                           & 19             & 13                & 3           & Transport            \\
FDP                         & 15             & 6                 & 9        & Policymakers               \\
Die Grünen                  & 14             & 8                 & 10       & Policymakers               \\
BDI                          & 11             & 7                 & 7       & Industry               \\
SPD                          & 10             & 7                 & 6                    & Policymakers   \\
DUH                          & 9              & 3                 & 5       & NGOs               \\
CDU                          & 8              & 4                 & 4                   & Policymakers    \\
Klimaallianz                 & 8              & 1                 & 8       & NGOs               \\
RWE                          & 8              & 3                 & 6       &    Electricity           \\
Westenergie                  & 8              & 1                 & 8       &    Electricity               \\
Zukunft Erdgas               & 8              & 3                 & 7           & Gas and heat             \\
BMVI                         & 7              & 6                 & 2                    & Policymakers   \\
Dena                         & 7              & 7                 & 4       & Research and think tanks               \\
FNB Gas                      & 7              & 5                 & 5               & Research and think tanks       \\
Fraunhofer ISE               & 7              & 4                 & 4                   & Research and think tanks   \\
Siemens                      & 7              & 3                 & 5               & Industry       \\
Universität Erlangen        & 7              & 3                 & 5                    & Research and think tanks  \\
Amprion                      & 6              & 3                 & 5                    &    Electricity  \\
BDEW                         & 6              & 3                 & 5                   &    Electricity   \\
CAR                          & 6              & 5                 & 2               & Research and think tanks       \\
DIHK                         & 6              & 1                 & 6           & Industry           \\
Daimler                      & 6              & 6                 & 4           & Transport           \\
IEA                          & 6              & 4                 & 5                    & Research and think tanks  \\
Salzgitter                   & 6              & 4                 & 4               & Industry       \\
Agora Energiewende           & 5              & 2                 & 5                    & Research and think tanks  \\
Bosch                        & 5              & 4                 & 3           & Transport           \\
DWV                          & 5              & 2                 & 5               & Industry       \\
Tennet                       & 5              & 1                 & 5              &    Electricity        \\
Toyota                       & 5              & 3                 & 4       & Transport               \\
Uniper                       & 5              & 4                 & 2               &    Electricity       \\
VDMA                         & 5              & 2                 & 4                   & Industry   \\
Viessmann                    & 5              & 1                 & 5               & Gas and heat       \\
Öko Institut                & 5              & 3                 & 4                    & Research and think tanks  \\
Airbus                       & 4              & 3                 & 3       & Transport               \\
BDH                          & 4              & 2                 & 4                    & Gas and heat    \\
DLR                          & 4              & 3                 & 4                    & Research and think tanks  \\
DVGW                         & 4              & 2                 & 4                     & Gas and heat   \\
EWE                          & 4              & 2                 & 4               &    Electricity       \\
GP Joule                     & 4              & 3                 & 3               & Electricity       \\
Hyundai                      & 4              & 3                 & 3           & Transport           \\
Nikola                       & 4              & 2                 & 4                   & Transport   \\
Open Grid Europe             & 4              & 3                 & 4                    & Gas and heat    \\
Prognos                      & 4              & 1                 & 4                    & Research and think tanks  \\
CEP                          & 3              & 1                 & 3                   & Industry   \\
E3G                          & 3              & 1                 & 3                    & Research and think tanks  \\
EWI                          & 3              & 1                 & 3                   & Research and think tanks   \\
Easac                        & 3              & 1                 & 3                   & Research and think tanks   \\
Gazprom                      & 3              & 1                 & 3               & Gas and heat         \\
Greenpeace                   & 3              & 1                 & 3       & NGOs               \\
H2 Mobility                  & 3              & 2                 & 2       & Transport               \\
Hydrogenious Technologies    & 3              & 2                 & 3               & Industry       \\
Monitoringkommission         & 3              & 1                 & 3               & Research and think tanks       \\
Senator of Hamburg           & 3              & 2                 & 3                   & Policymakers    \\
Shell Deutschland            & 3              & 1                 & 3                    & Gas and heat    \\
Siemens Energy               & 3              & 2                 & 2               &    Electricity       \\
TH Regensburg                & 3              & 2                 & 3           & Research and think tanks           \\
Thyssenkrupp                 & 3              & 3                 & 2               & Industry       \\
Total                        & 3              & 1                 & 3                    & Gas and heat    \\
VCI                          & 3              & 2                 & 3                   & Industry   \\
Vaillant                     & 3              & 1                 & 3                    & Gas and heat    \\
Wind2Gas                     & 3              & 2                 & 2                   & Industry   \\
Acatech                      & 2              & 2                 & 2               & Research and think tanks       \\
Avacon                       & 2              & 2                 & 1           &    Electricity           \\
BMZ                          & 2              & 1                 & 2                    & Policymakers   \\
Boston Consulting Group      & 2              & 1                 & 2                   & Research and think tanks   \\
Deutsche Reeder              & 2              & 1                 & 2           & Transport           \\
EEB                          & 2              & 1                 & 2           & NGOs           \\
ENBW                         & 2              & 1                 & 2                   &    Electricity   \\
Elring-Klinger                        & 2              & 1                 & 2          & Transport            \\
Energieagentur NRW           & 2              & 1                 & 2              & Research and think tanks        \\
Equinor                      & 2              & 1                 & 2                  & Gas and heat      \\
European Union               & 2              & 2                 & 2        &      European Union        \\
GE Power AG Mannheim         & 2              & 1                 & 2           &    Electricity           \\
GdW                          & 2              & 1                 & 2                    & Gas and heat    \\
Get H2 Nukleus               & 2              & 1                 & 2                     & Gas and heat   \\
Greenpeace Energy            & 2              & 2                 & 2                &    Electricity      \\
Horváth \& Partners         & 2              & 1                 & 2            & Research and think tanks          \\
HySolutions                  & 2              & 1                 & 2                  & Transport    \\
Hydrogen Council             & 2              & 1                 & 2               & Industry       \\
InnoEnergy                   & 2              & 1                 & 2              & Industry        \\
LNVG                         & 2              & 2                 & 2        & NGOs              \\
MCC                          & 2              & 1                 & 2           & Research and think tanks           \\
McKinsey                     & 2              & 2                 & 1                     & Research and think tanks \\
Northern Gas Networks        & 2              & 1                 & 2                   & Gas and heat     \\
Plastic Omnium               & 2              & 1                 & 2           & Transport           \\
RWTH Aachen                  & 2              & 1                 & 2                    & Research and think tanks  \\
Ruhr-Universität Bochum                        & 2              & 1                 & 2 & Research and think tanks                     \\
Shell                        & 2              & 2                 & 1                  & Gas and heat      \\
VDA                          & 2              & 1                 & 2           & Transport           \\
VKU                          & 2              & 1                 & 2                       & Gas and heat    \\
World Energy Council         & 2              & 2                 & 2       & Research and think tanks               \\
Wuppertal Institut           & 2              & 1                 & 2                    & Research and think tanks  \\
Air Liquide                  & 1              & 1                 & 1                    & Gas and heat    \\
Arcelor Mittal               & 1              & 1                 & 1                 & Industry     \\
Aurora Energy Research       & 1              & 1                 & 1                    & Research and think tanks  \\
BASF                         & 1              & 1                 & 1                   & Industry   \\
BDL                          & 1              & 1                 & 1               & Transport       \\
BeeZero                      & 1              & 1                 & 1                 & Transport     \\
Brot für die Welt           & 1              & 1                 & 1             & NGOs         \\
CAN                          & 1              & 1                 & 1               & NGOs       \\
Creavis                      & 1              & 1                 & 1               & Industry      \\
DNR                          & 1              & 1                 & 1                & NGOs      \\
Die Bahn                     & 1              & 1                 & 1               & Transport       \\
Die Linke                    & 1              & 1                 & 1                     & Policymakers  \\
Eon                          & 1              & 1                 & 1       &    Electricity               \\
FH Bergisch Gladbach         & 1              & 1                 & 1               & Research and think tanks       \\
FZ Jülich                   & 1              & 1                 & 1                & Research and think tanks      \\
First Berlin                 & 1              & 1                 & 1                   & Research and think tank   \\
Fraunhofer IMWS              & 1              & 1                 & 1                   & Research and think tanks   \\
Fraunhofer ISI               & 1              & 1                 & 1                    & Research and think tanks  \\
Fraunhofer IWU               & 1              & 1                 & 1                    & Research and think tanks  \\
HAW                          & 1              & 1                 & 1               & Research and think tanks       \\
HTKW                         & 1              & 1                 & 1                & Research and think tanks      \\
Hamburger Hochbahn           & 1              & 1                 & 1               & Transport       \\
Helmholtz-Institut                    & 1              & 1                 & 1               & Research and think tanks       \\
Hyundai Hydrogen Mobility AG & 1              & 1                 & 1               & Transport       \\
Hyundai Motors               & 1              & 1                 & 1       & Transport               \\
IASS                         & 1              & 1                 & 1       & Research and think tanks               \\
IfA                          & 1              & 1                 & 1               & Research and think tanks       \\
Ines                         & 1              & 1                 & 1              &Gas and heat        \\
LEE                          & 1              & 1                 & 1               & NGOs       \\
Mahle                        & 1              & 1                 & 1               & Transport       \\
NGO Kongo                    & 1              & 1                 & 1                  & NGOs    \\
NOW                          & 1              & 1                 & 1               & Transport       \\
Nowega                       & 1              & 1                 & 1                  & Gas and heat      \\
ÖNZ                         & 1              & 1                 & 1                & NGOs      \\
Roastom                      & 1              & 1                 & 1                  & Industry    \\
Schleswig-Holstein Netz AG                    & 1              & 1                 & 1                   & Electricity    \\
TU Chemnitz                  & 1              & 1                 & 1           & Research and think tanks           \\
Tesla                        & 1              & 1                 & 1              & Transport        \\
Universität Nürnberg       & 1              & 1                 & 1                 & Research and think tanks     \\
Vonovia                      & 1              & 1                 & 1        & Gas and heat                \\
WWF                          & 1              & 1                 & 1                   & NGOs   \\
\hline
Total                 & 614            & 339              & 424         &   \\ 
\hline
\end{xtabular}
\caption*{\footnotesize
For each actor, 
the table shows the number 
of coded storylines (Codes), 
of appearances in different newspaper articles (Articles),
of different coded storylines (Different SLs),
and its respective group (Group).}
}


\subsection{Temporal development of discourses}\label{appendix:att_storylines}

\begin{figure}[!htb]
\caption{Attention actors}\label{fig:att_actors}
\includegraphics[width=\textwidth]{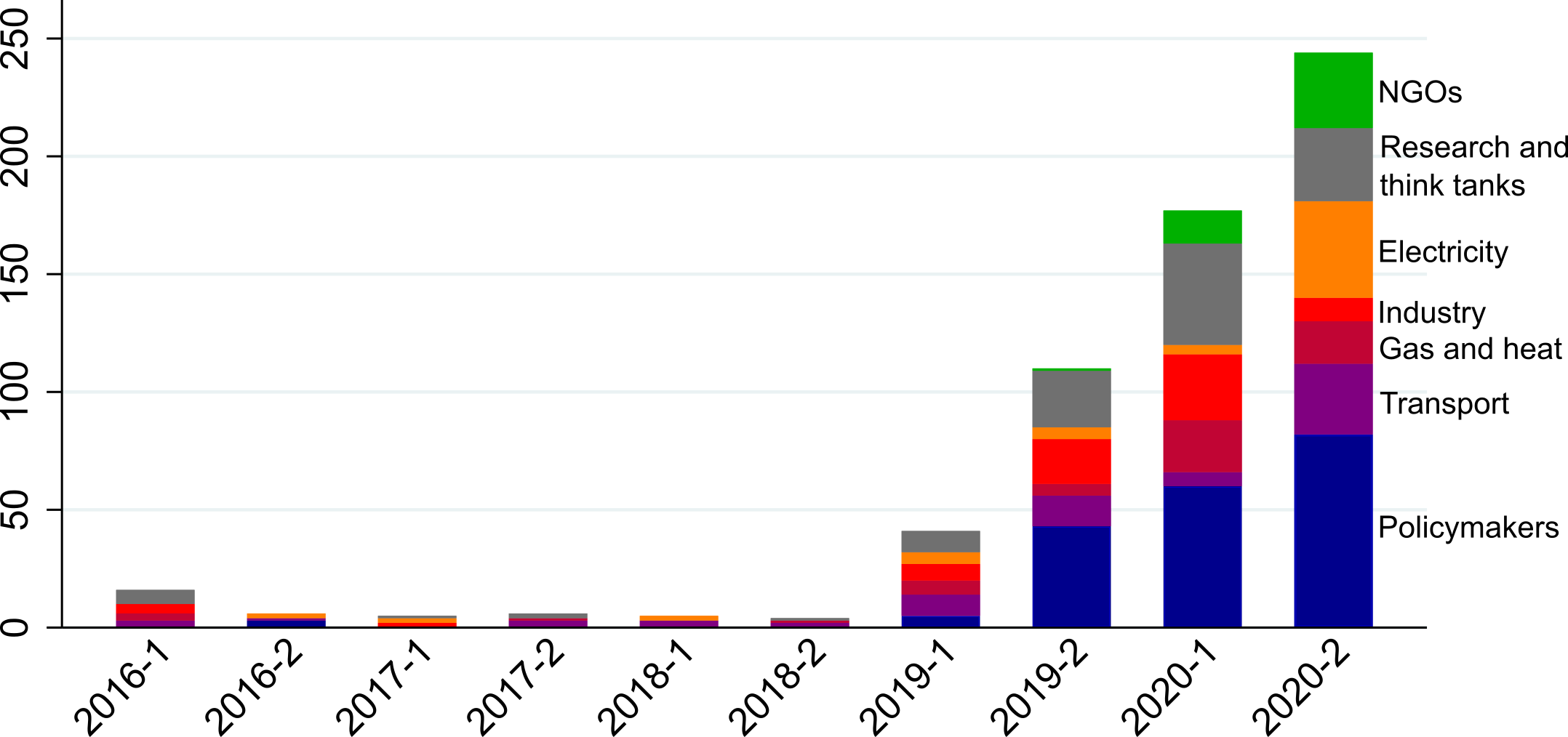}
\caption*{\footnotesize
The figure shows the frequency of coded storylines between 2016 and 2020 in half year bins stacked by actor groups.}
\end{figure}

\autoref{fig:att_actors} shows the participation of actor groups in the public debate over time.
Before 2019, 
storylines that characterize the ongoing discourses on hydrogen were scant. 
Newspaper articles covering hydrogen usually focused on fuel cell cars, 
or occasionally on power-to-X. 
The ongoing discourses on hydrogen started in 2019. 
The start coincides with the declaration by the German chancellor Angela Merkel that net-zero emissions by 2050 would be the new climate target. 
The number of of coded storylines tripled between the first and the second half of 2019. 
In 2019, the public debate was overall dominated by policymakers, research institutes and think tanks,
and industry actors.
In 2020, the absolute number of codes for these groups remained stable, 
but additional actor groups entered the public debate: 
the gas and heat sector became increasingly visible in the first half of 2020,
while NGOs, the electricity and transport sector, and the EU joined in the second half.

\begin{figure}[!htb]
\caption{Attention storylines}\label{fig:att_storylines}
\includegraphics[width=\textwidth]{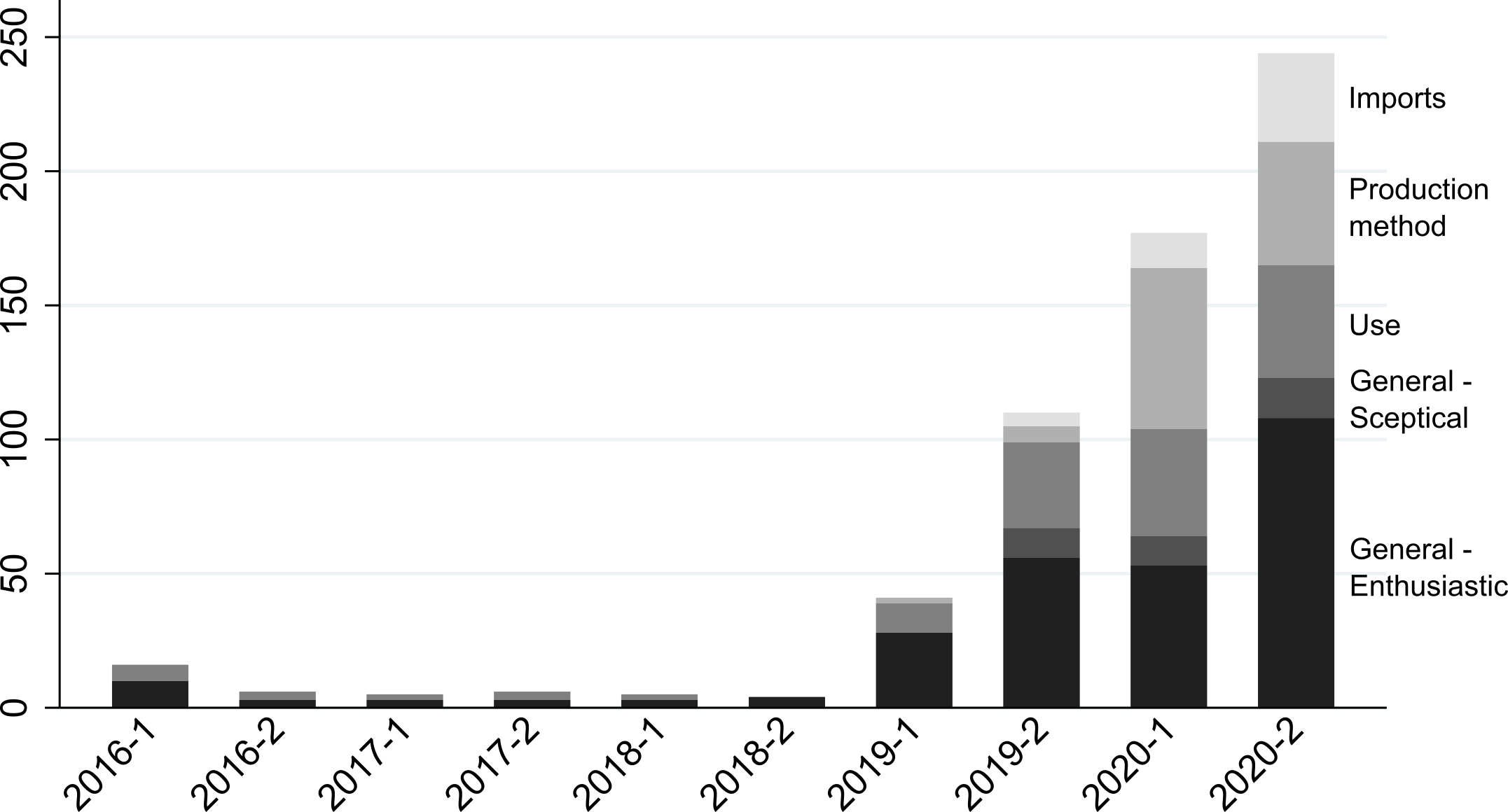}
\caption*{\footnotesize\textit{Notes:}
The figure shows the frequency of coded storylines between 2016 and 2020 in half year bins stacked by storylines.}
\end{figure}

\autoref{fig:att_storylines} shows how the three conflicts, as well as skeptical and enthusiastic storylines in the general discussion on hydrogen evolved over time. 
In 2019,
the discourse is dominated by enthusiastic storylines and the conflict about its use. 
The conflict about production methods becomes more important in early 2020, 
while discussion about imports begin only in late 2020. 
skeptical storylines appear in late 2019, but overall remain an exception.

\end{document}